# JoinActors: A Modular Library for Actors with Join Patterns

Ayman Hussein[a], Philipp Haller[b], Ioannis Karras[a], Hernán Melgratti[c], Alceste Scalas[a], and Emilio Tuosto[d]

a Technical University of Denmark, Denmark
b KTH Royal Institute of Technology, Sweden
c University of Buenos Aires and Conicet, Argentina
d Gran Sasso Science Institute, Italy

**Abstract** *Join patterns* are a high-level programming construct for message-passing applications. They offer an intuitive and declarative approach for specifying how concurrent and distributed components coordinate, possibly depending on complex conditions over combinations of messages. Join patterns have inspired many implementations — but most of them are not available as libraries: rather, they are domain-specific languages that can be hard to integrate into pre-existing ecosystems. Moreover, all implementations ship with a predefined matching algorithm, which may not be optimal depending on the application requirements. These limitations are addressed by JoinActors, a recently published library which integrates join patterns in the off-the-shelf Scala 3 programming language, and is designed to be modular w.r.t. the matching algorithm in use.

In this work we address the problem of designing, developing, and evaluating a modular join pattern matching toolkit that (1) can be used as a regular library with a developer-friendly syntax within a pre-existing programming language, and (2) has an extensible design that supports the use and comparison of different matching algorithms.

We analyse how JoinActors achieves goals (1) and (2) above. The paper that introduced JoinActors only briefly outlined its design and implementation (as its main goal was formalising its novel *fair matching semantics*). In this work we present and discuss in detail an improved version of JoinActors, focusing on its use of metaprogramming (which enables an intuitive API resembling standard pattern matching) and on its modular design. We show how this enables the integration of multiple matching algorithms with different optimisations and we evaluate their performance via benchmarks covering different workloads.

We illustrate a sophisticated use of Scala 3's metaprogramming for the integration of an advanced concurrent programming construct within a pre-existing language. In addition, we discuss the insights and "lessons learned" in optimising join pattern matching, and how they are facilitated by JoinActors's modularity — which allows for the systematic comparison of multiple matching algorithm implementations.

We adopt the *fair join pattern matching* semantics and the benchmark suite from the paper that originally introduced JoinActors. Through extensive testing we ensure that our new optimised matching algorithms produce exactly the same matches as the original JoinActors library, while achieving significantly better performance. The improved version of JoinActors is the companion artifact of this paper.

This work showcases the expressiveness, effectiveness, and usability of join patterns for implementing complex coordination patterns in distributed message-passing systems, within a pre-existing language. It also demonstrates promising performance results, with significant improvements over previous work. Besides the practical promise, JoinActors's modular design offers a research playground for exploring and comparing new join pattern matching algorithms, possibly based on entirely different semantics.

# The Art, Science, and Engineering of Programming

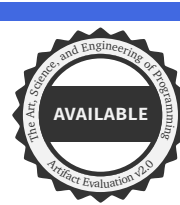
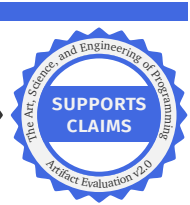

**JoinActors: A Modular Library for Actors with Join Patterns**

■ **Listing 1** A payment service implemented with the JoinActors library.

```scala
1  …
2  type PaymentEvent = PaymentRequested | MerchantValidated | CustomerValidated
3
4  Actor[PaymentEvent | Shutdown, Unit]{
5      receive { self => { // 'self' is the actor "address", usable for sending messages
6          case PaymentRequested(id1) &:& MerchantValidated(id2) &:& CustomerValidated(id3)
7              if (id1 == id2) && (id2 == id3) =>
8              coreService ! PaymentSucceeded(id1)
9              log(s"Payment service handled payment request $id1")
10             Continue
11
12         case Shutdown() => Stop(()) }
13     }(matcherAlgorithm) // Select the join pattern matching algorithm to use
14 }
```

## 1  Introduction

Many software applications are based on concurrent and distributed components that exchange data and coordinate via message passing. This kind of interaction can be found, e.g., in cloud microservices, edge computing meshes, and Internet of Things (IoT) deployments. These applications often need to perform specific tasks only once a certain set of messages has been received, and certain predicates over their payloads are satisfied. This may require the development of non-trivial bookkeeping logic to keep track of the observed messages, resulting in verbose and and error-prone code.

*Join patterns* provide a programming abstraction that addresses these coordination challenges. The idea was first introduced in the join calculus [13], and it extends the familiar notion of pattern matching to message-passing settings: join patterns match against sets of messages and "fire" once all required messages are received. Join patterns may also incorporate *guards* (i.e., predicates over message payloads) that must hold in order to fire the pattern. By capturing coordination logic in concise, declarative specifications, join patterns can reduce reliance on manual state management and improve the clarity and reliability of distributed applications [29, Section 4].

A recent implementation of join patterns is the JoinActors library [17] for Scala 3. It offers a programming interface inspired by well-known actor [1] frameworks such as Akka and Pekko[1] — except that JoinActors allows actors to react to combinations of messages (represented as join patterns) rather than individual messages.

**Example: a payment service using the JoinActors library**   Listing 1 shows a simplified mobile payment back-end service. The service is implemented as an actor that receives messages in its *mailbox* and processes them via join pattern matching. The actor constructor (from line 4) says that the actor accepts messages of type **PaymentEvent** (defined on line 2) or **Shutdown**,[2] and returns a result of type **Unit** whenever it terminates. The actor's behaviour is defined in the receive block (lines 5–13), where two join patterns (on lines 6–7 and 12) are fired whenever the mailbox receives a corresponding combination of messages.

---
[1] https://akka.io/, https://pekko.apache.org/ (Visited on 2026-02-12).

The join pattern on lines 6–7 specifies two conditions:

1. The mailbox must contain three messages of types **PaymentRequested**, **MerchantValidated**, and **CustomerValidated**, respectively. Such messages are sent by other agents participating in the payment workflow, and they can be delivered in any order; and
2. The *guard* (i.e., the **if** condition on line 7) must evaluate to **true** for those three messages. This happens when their payloads (captured by variables id1, id2, and id3) are equal, meaning that they refer to the same payment identifier.

Whenever these two conditions are satisfied, the join pattern is fired: i.e., the three matching messages are removed from the mailbox, and the code following **=>** (lines 8–10) is executed. When this happens, the actor sends a **PaymentSucceeded** message to the core service (line 8) and logs the event (line 9). The final **Continue** (line 10) indicates that the actor remains active and available for subsequent message processing.

The join pattern on line 12, instead, fires whenever a message of type **Shutdown** is received. When this happens, **Stop**(()) causes the actor to stop executing and return the unit value ().

Observe that without join patterns, the payment service would need to implement manual bookkeeping logic to check whether a newly-arrived message satisfies the guard when combined with previously-received messages — and therefore, such messages would need to be maintained in an internal state. This would lead to more complex and error-prone code, especially when multiple join patterns overlap and "compete" on the same messages (as shown in the extended example in Section 3.2).

Finally, on line 13 of Listing 1, the parameter matcherAlgorithm determines how messages are matched against the declared join patterns. This is a crucial element in the design of the JoinActors library: it allows for different matching algorithms to be selected, depending on the application's requirements.

**Contribution: a modular, optimised, and extensible library for join pattern matching** In this work, we present a revised and extended version of JoinActors which significantly improves the modularity and performance w.r.t. [17] (which only presented an initial proof-of-concept implementation of the library and a high-level outline of its design). We illustrate how JoinActors leverages Scala 3's metaprogramming features to embed join pattern matching into a preexisting language, while providing a developer-friendly API. We also leverage JoinActors's modular architecture to implement and systematically compare different optimised matching algorithms which realise the *fair matching semantics* in [17]; we show how their optimisations significantly affect the matching performance on different workloads.

---

[2] In Listing 1, the symbol **|** denotes a type union in Scala 3.

**JoinActors: A Modular Library for Actors with Join Patterns**

We provide some relevant background in Section 2. After an overview of JoinActors (Section 3), we present its design and implementation in detail (Section 4), focusing on its modularity and metaprogramming aspects, and on its optimised matching algorithms; then, we evaluate the algorithms' performance (Section 5). We discuss related work in Section 6 and conclude in Section 7.

## 2 Background

This section provides background information on the key concepts behind our JoinActors library: the actor model (Section 2.1), fair join pattern matching (Section 2.2), and metaprogramming in Scala 3 (Section 2.3).

### 2.1 Actor Model

The JoinActors API implements a variant of the popular *actor model* [1, 18]. In this model, computation is structured around actors — independent entities that encapsulate private state and process messages sequentially. Actors interact exclusively via asynchronous message passing using non-blocking message sending; after receiving a message, an actor may update its internal state (e.g. to change its own behaviour for processing further messages), send messages to other actors, and spawn new actors. Actors can exchange their addresses in messages, thus enabling the dynamic reconfiguration of the actor network at runtime. In traditional actor systems such as Akka/Pekko, the actors react to each incoming message. In contrast, our JoinActors library allows actors to react to combinations of messages from their mailboxes; such combinations are represented as join patterns as outlined in Listing 1.

### 2.2 Fair Join Pattern Matching

Most implementation of join patterns in the literature use a non-deterministic matching semantics: if multiple combinations of messages match a join pattern, one is selected arbitrarily [16, 29]. This, in principle, means that some messages in the mailbox may never be selected despite "being matchable". The recent work [17] introduces a *deterministic* join pattern matching semantics based on the concept of *fair join pattern matching*: when a join pattern matches multiple message combinations, the match consuming the oldest messages is prioritised — and as a result, matchable messages will not be left indefinitely in the mailbox. The early version of JoinActors introduced in [17] implements this fair matching semantics with two algorithms:

**Stateless "brute force" matching (BruteForceMatcher).** This algorithm waits until the mailbox contains enough messages to "fill" the join pattern. Then, it filters messages by type, enumerates all message combinations leading to a match, and selects the fairest one. This algorithm directly implements the formal fair matching semantics, but it re-computes prior non-matching message combinations whenever a new message arrives, causing inefficiency depending on the workload.

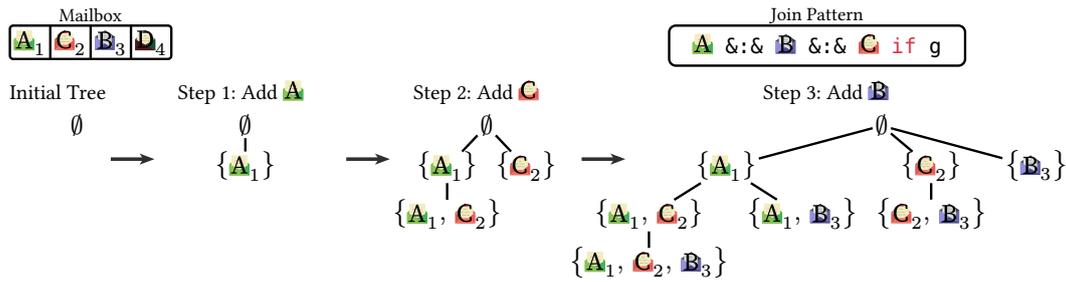

**Figure 1** Incremental ramification performed by the **StatefulTreeMatcher** algorithm in [17], described in Example 2.1.

**Stateful matching (StatefulTreeMatcher).** This algorithm maintains a state, as a *matching tree* that tracks which previously-received messages partially complete a join pattern. When a new message arrives, the algorithm checks the matching tree and avoids recomputing prior non-matching combinations. The matching tree contains partial matches and is constructed incrementally ("ramified") as new messages arrive. Conceptually, this is reminiscent of RETE-style [11] forward chaining for rule-based AI systems: join patters are akin to RETE *productions*, while new messages reaching the mailbox are akin to new data being added to the RETE working memory; the nodes of the matching tree represent progressively stronger tests that filter messages and cache intermediate results until a terminal node is reached (where a set of messages completes the pattern/production). However, unlike RETE-style algorithms, the ramification process maintains an invariant: a depth-first traversal of the matching tree always yields the fairest possible match (if any). The tree is pruned whenever a message completes the pattern, in different ways depending on whether the guard evaluates **true** or **false**. This algorithm usually has better performance than the stateless one. The ramification process is described in Example 2.1 below.

**Example 2.1** (Ramification of a matching tree)**.** Figure 1 illustrates how the **StatefulTreeMatcher** algorithm in [17] progressively matches 4 incoming messages (mailbox in the top-left corner) against the join pattern on the top-right corner. The matching tree is initially empty, and is ramified as new messages arrive: e.g., the tree at step 2 says that the pattern is partially matched by message **A** (at position 1 in the mailbox) alone, or by **C** (at position 2) alone, or by **A** and **C** together. At step 3, a depth-first traveral leads to a leaf with a potential complete match that uses messages **A**, **C**, and **B**. This potential match is used to evaluate the guard $g$: if the result is **true**, the match succeeds, the messages are removed from the mailbox, and all subtrees descending from them are pruned; otherwise, only the leaf is pruned and never checked again, as it represents a failed match. ◾

The paper [17] also introduces a benchmarking suite for join pattern matching, that we adopt for our evaluation in Section 5.

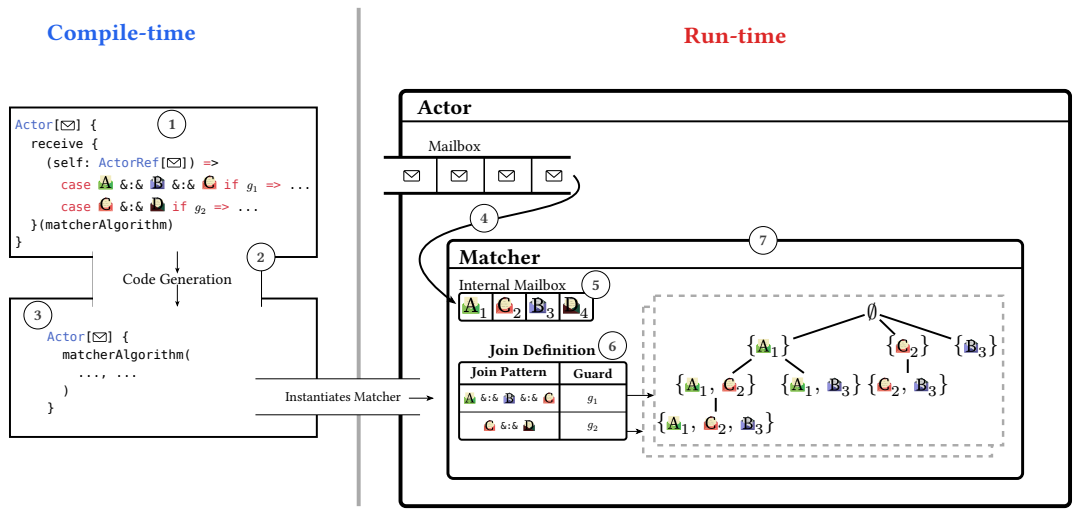

**Figure 2** Overview of the JoinActors library operation at compile-time and run-time.

## 2.3 Metaprogramming in Scala 3

Scala 3 supports the definition of functions and arguments that are inlined in their call site at compile time, via the inline modifier. Inline functions can also be used to analyse, construct, and return code fragments at compile time, thus becoming *macros* for code generation [34]. For the analysis and construction of code fragments, Scala 3 provides both a quote-and-splice mechanism [32] and a reflection API [33]: the former is type-safe but somewhat limited in its expressiveness, whereas the latter has weaker type checks (i.e., it potentially allows for generating ill-typed code) but offers a more fine-grained control over code inspection and transformation. Internally, JoinActors uses both mechanisms, although in Section 4.2.1 we will mainly exemplify its use of the reflection API.

## 3 Overview of the JoinActors Library and Its Usage

We now provide a more detailed overview of how JoinActors operates at compile-time and run-time (Section 3.1), and an extended example of its usage (Section 3.2).

### 3.1 JoinActors at Compile-Time and Run-Time

The overall architecture and usage of the JoinActors library is illustrated in Figure 2:

1. The programmer using JoinActors writes an **Actor** declaring the type of messages it accepts (similarly to Listing 1). In Figure 2, this type is depicted as a closed envelope. The body of the actor is specified with the receive macro (itself provided by JoinActors, and discussed in Section 4.2.1), whose first argument is expected to be the code of a *partial function*, i.e., a **match** expression whose cases define the join patterns that the actor will handle. In Figure 2, the clauses of the join pattern are depicted as opened envelopes, as the join pattern can inspect the specific type and payload of each message.

2. At compile time, the receive macro inspects the **match** expression given as argument, and extracts the information needed to generate join pattern matching code.

3. After code generation, the body of the actor is replaced by code that instantiates the matcherAlgorithm (the second argument provided to receive at step 1) with the join pattern matching information extracted at step 2. (More details in Section 4.)

4. At runtime, the actor retrieves messages from its mailbox one at a time and forwards them to a **Matcher** object instantiated by the matcherAlgorithm.

5. The **Matcher** object handles the received messages, storing them internally. In Figure 2, these messages are shown as opened envelopes, reflecting that both the types and values of the messages are now fully available; their numbers indicate the order in which they were received, which is relevant for *fair matching* (Section 2.2).

6. The **Matcher** object then evaluates the stored messages against the user-supplied join pattern maching definitions.

7. The selected matcherAlgorithm determines how the **Matcher** object works. Figure 2 illustrates the case where the **Matcher** is an instance of the **StatefulTreeMatcher** class and maintains a matching tree for each join pattern (as in Example 2.1).

### 3.2 Example: Factory Shop Floor Monitoring

We now illustrate the use of JoinActors in an application with a significant number of non-trivial join patterns. Listing 2 shows the (simplified) code of a factory shop floor monitoring application (an extension of the opening example in [17]). The application receives messages describing relevant **Event**s on the shop floor (line 1):

1. A machine reports a **Fault** to request maintenance. **Fault** messages include the machine ID, a unique request ID, and a timestamp;

2. A worker **Fix**es a fault. **Fix** messages include the machine ID, request ID, and a timestamp;

3. Periodic **Mark** messages that signal the passage of time, and are used to determine whether maintenance requests have been addressed within a desired time frame. **Mark**s include a timestamp and a set of fault IDs tracked by the mark.

The application detects combinations of messages reaching its mailbox, and reports anomalies or quality-of-service violations. To this end, the application sometimes produces and sends to its own mailbox some **RecentFix** messages — which are similar to **Fix** events and maintain information about recently fixed faults.

The join pattern matching support provided by JoinActors allows for declaratively specifying such message combinations, as shown in Listing 2:

**Rapid Recurrence of Faults:** The first join pattern (lines 7–9) matches a sequence of messages denoting that a machine has been recently fixed, but the same machine subsequently failed again within 30 minutes.

**Delayed Maintenance:** The second and third join patterns (lines 14–22) capture scenarios where a fault is reported but not fixed within 10 minutes. If a worker only

# JoinActors: A Modular Library for Actors with Join Patterns

▎ **Listing 2** A simplified factory monitoring application using the JoinActors library. The application is described in Section 3.2.

```
1  type Event = Fault | Fix | Mark | ...
2
3  def monitor(matcherAlgorithm: MatcherFactory) = Actor[Event | RecentFix | Shutdown, Unit] {
4      receive { self => {
5
6          // Rapid reoccurrence of faults
7          case RecentFix(mid1, rid1, ts1) &:& Fault(mid2, rid2, ts2)
8                      if mid1 == mid2 && rid1 != rid2 && ts2 >= ts1 && ts2 - ts1 < THIRTY_MIN =>
9              self ! Fault(mid3, rid3, ts3) // Re-enqueue latest fault
10             println(s"Machine ${mid1} broke within 30 minutes after maintenance!")
11             Continue
12
13         // Delayed maintenance
14         case Fault(_, rid1, ts1) &:& Fix(mid, rid2, ts2) if rid1 == rid2 && ts2 - ts1 >= TEN_MIN =>
15             println(s"Fault ${rid1} only fixed after ${(ts2 - ts1) / ONE_MIN} minutes!")
16             self ! RecentFix(mid2, rid2, ts2) // Self-notify that a fault was recently fixed
17             Continue
18         case Fault(mid1, rid1, ts1) &:& Mark(ts2, tracked)
19          if ts2 - ts1 >= TEN_MIN && !tracked.contains(rid1) =>
20             self ! Fault(mid1, rid1, ts1) // Re-enqueue the unhandled fault
21             self ! Mark(ts2, tracked + rid1) // Re-enqueue the mark, tracking the unhandled fault
22             println(s"Fault ${rid1} not taken for ${(ts2 - ts1) / ONE_MIN} minutes!")
23             Continue
24
25         // Timely maintenance
26         case Fault(_, rid1, ts1) &:& Fix(mid2, rid2, ts2) if rid1 == rid2 && ts2 - ts1 < TEN_MIN =>.
27             self ! RecentFix(mid2, rid2, ts2) // Self-notify that a fault was recently fixed
28             Continue // Fault handled within 10 minutes, nothing to report
29
30         // Disposal of old marks and recent fixes after some time
31         case Mark(ts1, _) &:& Mark(ts2, tracked2) if ts2 - ts1 >= 2 * TEN_MIN =>
32             self ! Mark(ts2, tracked2) // Re-enqueue newest mark only
33             Continue
34         case RecentFix(_, _, ts1) &:& Mark(ts2, tracked2) if ts2 - ts1 >= 4 * TEN_MIN =>
35             self ! Mark(ts2, tracked2) // Re-enqueue the mark only
36             Continue
37
38         case Shutdown() => Stop(())
39      }
40    }(matcherAlgorithm)
41 }
```

fixes the fault after this threshold, or if the fault has been unhandled for more than 10 minutes (according to a **Mark** message), then the application reports the quality-of-service violation. To avoid repeated reports involving the same fault and the same **Mark**, the guard of the second join pattern (line 18) requires that the fault was not already tracked in the **Mark** message; then, the application re-enqueues a modified **Mark** that tracks the fault (line 20).

**Timely Maintenance:** The fourth join pattern (lines 25-27) simply records a **RecentFix** when a fault is fixed within 10 minutes.

**Obsolete Events Cleanup:** The fifth and sixth join patterns (lines 30-35) discard **Mark** and **RecentFix** events after some time; this is checked whenever a new mark is received.

Finally, the last join pattern (line 37) stops the monitor when it receives a **Shutdown** message.

Observe that a conventional actor-based implementation of Listing 2 would need to process each incoming message individually, and would include dedicated code for tracking message combinations — including possible out-of-order deliveries (e.g., a **Fix** may be observed before the corresponding **Fault**). This would significantly increase code complexity and the risk of errors. Moreover, the patterns in Listing 2 often use the same message types, hence the arrival of a message may activate multiple patterns: in this case, JoinActors deterministically selects the *fairest* match that consumes the oldest messages [17]. In this example, this is useful e.g. to ensure that the oldest **Mark** and **RecentFix** events are disposed of in lines 30–35.

## 4 Design and Implementation of JoinActors

We now present the design (Section 4.1) and implementation (Section 4.2) of the JoinActors library. We first highlight how the design decisions contribute to the library's functionality and extensibility. We then describe the implementation, emphasizing the use of Scala 3's metaprogramming facilities.

### 4.1 Design and Extensibility

The API of the JoinActors library is designed for usability and expressiveness: programmers specify join patterns using a syntax resembling standard Scala 3 pattern matching, as shown in Listing 1. This is achieved by the receive macro, which requires its first argument to be a well-typed Scala 3 (partial) function with a **match** expression.

Listing 3 shows the general structure of an actor using the JoinActors API. In the constructor **Actor**[**M**,**T**], the type parameter **M** denotes the type of messages that the actor may receive, while **T** denotes the actor's return type.[3]

---

[3] More precisely: spawning an actor creates a **Future**[**T**]; whenever the actor terminates, it completes that future with a value of type **T**.

▌ **Listing 3** General structure of an actor implemented with the JoinActors library.

```
1  Actor[M, T] {
2      receive { (self : ActorRef[M]) => {
3          case P₁ if g₁ => e₁
4          case P₂ if g₂ => e₂
5          ...
6          case Pₙ if gₙ => eₙ
7      }(matcherAlgorithm)
8      }
9  }
```

▌ **Listing 4** The type signature of the receive macro, using the traits in Listing 5.

```
1  inline def receive[M, T] (inline closure: ActorRef[M] => PartialFunction[Any, Result[T]])
2                           (inline matcherAlgorithm: MatcherFactory) :  Matcher[M, Result[T]]
```

The constructor **Actor[M,T]** takes one argument: an instance of the receive macro wrapping a block of code that, given the actor's self reference (of type **ActorRef[M]**), specifies the join patterns to be matched. In Listing 3, each $P_i$ denotes a join pattern over messages of type **M**, possibly returning a value of type **T**. The message type **M** is typically defined as a type union, or a **sealed** family of **case class**es or **enum** variants;[4] moreover, each $P_i$ is a "conjunction" of messages of the form $M_1$(...) **&:&** $M_2$(...) **&:&** ... **&:&** $M_m$(...) (as shown in Listing 1), where each $M_j$ (for $j \in 1..m$) is a message type (usually a subtype of **M**), and the operator **&:&** is provided by JoinActors. Note that the number of messages in the conjunction may vary between patterns. Each $g_i$ is a *guard*, i.e., a boolean expression that may reference variables bound in the corresponding join pattern $P_i$. Each $e_i$ is an expressions that is executed when the corresponding join pattern and guard are satisfied; each $e_i$ must return either **Continue** (to keep executing the actor) or **Stop**(result), where result is an expression having the actor's return type **T**.

The second argument to receive, i.e. matcherAlgorithm, specifies the join pattern matching algorithm. This argument enforces a uniform instantiation mechanism for different algorithms, which are required to implement the **Matcher** trait.

More precisely, the type signature of the receive macro is shown in Listing 4. Its first argument is the (code of the) partial function outlined in Listing 3, which defines the **case** clauses with join patterns, guards, and expressions. Its second argument is an instance of the **MatcherFactory** trait (discussed in Listing 5 below) which instantiates the chosen matching algorithm. When the receive macro is executed, it extracts join pattern information from its first argument and passes it to the matcherAlgorithm to instantiate the actual **Matcher** (more details in Section 4.2.1).

---

[4] Scala 3 enums encode finite alternatives (sum types), while case classes define product types; these are usually combined under a sealed trait to provide closed sums.

▪ **Listing 5** The **Matcher** trait for join pattern matchers, and the **MatcherFactory** trait for constructing them.

```
1  trait Matcher[M, T]:
2      def apply(mbox: Mailbox[M])(self: ActorRef[M]): T
3
4  trait MatcherFactory:
5      def apply[M, T]: List[JoinPattern[M, T]] => Matcher[M, T]
```

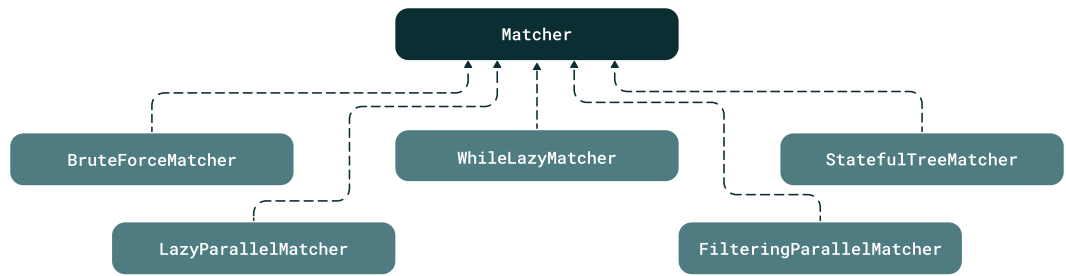

▪ **Figure 3** Overview of some of the join pattern matching algorithms provided by JoinActors.

**Extending JoinActors with new matching algorithms** We designed the matching algorithms' interface with the aim of making the JoinActors library modular and extensible. Figure 3 provides an overview of some of the matching algorithms already available in JoinActors (and discussed in Section 4.2.2).

To provide a new matching algorithm, a programmer is expected to:

1. Define a class with the algorithm logic, implementing the **Matcher** trait (Listing 5);
2. Create a companion object responsible for instantiating the new matcher class, by extending the **MatcherFactory** trait (Listing 5). This companion object can then be passed as second argument to the receive macro (see Listing 4).

As shown in Listing 5, a **Matcher** operates by being applied to a mailbox mbox (which is a thread-safe blocking queue of incoming messages) with the actor's own "address" self; when a match occurs, the matcher is expected to execute the corresponding code (i.e., $e_i$ in Listing 3) and return its result (of type **T**).

## 4.2 Implementation

We now describe the implementation of the JoinActors library. We first explain the code generation process (Section 4.2.1), then we present some of the join pattern matching algorithms included in the library, and their optimisations (Section 4.2.2).

### 4.2.1 Code Generation Macros

The receive macro inspects the typed AST of the inlined closure that it receives as first argument (having type **ActorRef**[**M**] => **PartialFunction**[**Any**, **Result**[**T**]], as per Listing 4). This inspection enables the extraction of join patterns, guards, and right-hand side expressions for each case, as outlined in Listing 6. The body of the closure is expected

- **Listing 6** Simplified fragment of the receive macro that inspects the code of the closure passed as first argument (as per signature in Listing 4) using the Scala 3 reflection API [33]. On line 4, self is the closure argument (as in Listing 3), while body is the body of the closure. On line 6, cases is the list of pattern matching **case**s in the body of the closure; each one is decomposed to extract the case pattern, guard, and right-hand-side (corresponding to $P_i$, $g_i$, and $e_i$ in Listing 3, respectively); which are used on line 9 to generate code representing a join pattern.

```scala
closureAST match
  case Inlined(_, _, Inlined(_, _, Block(_, Block(List(stmt), _)))) =>
    stmt match
      case DefDef(_, List(TermParamClause(List( self ))), _, Some(Block(_, Block(List( body ), _)))) =>
        body match
          case DefDef(_, _, _, Some(Match(_, cases ))) =>
            cases.flatMap {
              case CaseDef( pattern, guard, rhs ) =>
                generateJoinPattern[M, T](pattern, guard, rhs, self)
            }
```

- **Listing 7** The data type representing a join pattern. **LookupEnv** is an environment that maps the variables occurring in a message pattern to their values.

```scala
case class JoinPattern[M, T](
    guard: LookupEnv => Boolean, // Closure for evaluating the guard
    rhs: (LookupEnv, ActorRef[M]) => T, // Closure for executing the RHS of this join pattern
    size: Int, // Number of messages needed to match this join pattern
    matcherUtils: MatcherUtils[M] // Utilities for matcher implementations (e.g. for computing LookupEnv)
)
```

to be a **Match** instance (lines 6–9 in Listing 6), i.e., a Scala 3 AST node representing a pattern matching construct. Each **case** of the pattern matching construct contains three components: the message pattern, an optional guard, and the right-hand side (corresponding to $P_i$, $g_i$, and $e_i$ in Listing 3, respectively).

As mentioned in Section 4.1, the message pattern has the form $M_1$(...) **&:&** $M_2$(...) **&:&** ... **&:&** $M_m$(...), where each $M_j$ (for $j \in 1..m$) is a message type. Hence, the receive macro needs to extract the message types from the pattern, including the types of the message payloads. This extraction is performed by generateJoinPattern (line 9 in Listing 6), which generates code that instantiates the **JoinPattern** class shown in Listing 7 — where, the field **MatcherUtils**[M] in Listing 7 encapsulates a suite of utilities to assist the matching algorithms. The shape of the generated code is outlined in Example 4.1 below – after which we further elaborate on the implementation of generateJoinPattern.

**Example 4.1** (Code generation for a join pattern). Listing 8 shows the code generated for the first join pattern in the payment service example (Listing 1). Both the guard and rhs functions take a **LookupEnv** instance as an argument, which maps payload identifiers to the values extracted from the selected mailbox message. The guard is

■ **Listing 8**  The translation of the first join pattern in paragraph 1 after the execution of the receive macro (slightly simplified for readability).

```
JoinPattern[PaymentEvent, Unit](
    guard = (env: LookupEnv) =>
                (env("id1").asInstanceOf[Int] == env("id2").asInstanceOf[Int]) &&
                (env("id2").asInstanceOf[Int] == env("id3").asInstanceOf[Int]),
    rhs = (env: LookupEnv, self: ActorRef[PaymentEvent]) => {
                coreService ! PaymentSucceeded(env("id1").asInstanceOf[Int])
                log(s"Payment service handled payment request ${env("id1").asInstanceOf[Int]}")
                Continue
    },
    size = 3,
    matcherUtils = MatcherUtils[M](...)
)
```

evaluated in this environment to determine whether the join pattern matches the selected messages: observe that the code of the guard in Listing 1 has been rewritten to use the **LookupEnv** instance env, so that, e.g., the variable id1 in Listing 1 is now accessed as env("id1").**asInstanceOf**[**Int**].[5] When the guard evaluates to **true**, the rhs is invoked with the same **LookupEnv** used for the guard to execute the computation specified by the right-hand-side of the join pattern; again, notice that the code of the rhs has been rewritten similarly to the guard.    ◾

**Generating a JoinPattern instance via generateJoinPattern**    The function generateJoinPattern (line 9 in Listing 6) is responsible for the code generation outlined in Example 4.1 and Listing 8. For each **CaseDef** extracted from the code passed to the receive macro, generateJoinPattern performs further analysis to determine whether the **CaseDef** denotes a unary message pattern (e.g., **case** **A**(x) **=>** ...) or an *n*-ary pattern composed with the **&:&** operator (e.g., **case** **A**(x) **&:&** **B**(y) **=>** ...). In both cases, it delegates to specialised generators that recover message and payload type information, track the variables bound by the pattern, and synthesise utility functions to perform run-time message matching. Such functions are assembled into an instance of **MatcherUtils**[**M**] (line 5 of Listing 7 and line 11 of Listing 8) that matching algorithms can use to process incoming messages of generic type **M**. An instance of **MatcherUtils**[**M**] has two key components:

1. a mapping that groups pattern positions by message type; and

---

[5] The downcast **asInstanceOf**[**Int**] is needed because the **LookupEnv** maps bound variables to values of type **Any**. This is because message payloads (and thus, the variables bound by message patterns) can be of different types, hence **LookupEnv** resorts to **Any** as their only guaranteed supertype. The receive macro infers the downcast type while inspecting the AST of the join pattern: e.g., env("id1") is downcast to **Int** because in Listing 1 the variable id1 is bound by the message pattern **PaymentRequested**(id1), whose payload has type **Int**.

■ **Table 1** Optimisations implemented by the join pattern matching algorithms in Figure 3.

| Matcher Type | Optimisation Description |
| --- | --- |
| **BruteForceMatcher** | Cache-efficient data structures |
| **StatefulTreeMatcher** | Cache-efficient data structures |
| **WhileLazyMatcher** | Cache-efficient data structures, efficient iteration, lazy ramification |
| **LazyParallelMatcher** | Cache-efficient data structures, efficient iteration, lazy ramification, parallel ramification |
| **FilteringParallelMatcher** | Cache-efficient data structures, efficient iteration, lazy ramification, parallel ramification, partial evaluation of guards |

2. a mapping that associates each pattern position with *(i)* a function for recognising the exact type of an incoming message, and *(ii)* a function that populates a **LookupEnv** with the payload values extracted from the incoming message.[6]

This organization allows matcher implementations to first narrow down which pattern position(s) may fit an incoming message (depending on its type), and then extract payload values.

### 4.2.2 Join Pattern Matching Algorithms and Optimisations

We now reprise the join pattern matching algorithms listed in Figure 3 to discuss the optimisations they implement w.r.t. [17], which only included early versions of the **BruteForceMatcher** and **StatefulTreeMatcher** (as discussed in Section 2.2). These optimisations are summarised in Table 1, and described below; further details are available in [22]. Notably, the modular design of JoinActors allows us to implement and evaluate these optimisations incrementally. We compare the different algorithms in Section 5, highlighting how different optimisations yield different perfomance depending on the workload.

**Cache-efficient data structures** The early version of JoinActors in [17] uses immutable linked-list collections (**List**, **Queue**). This design incurred $O(n)$ random access, poor cache locality, and boxing overhead. We replaced these structures with **ArraySeq**s, which offer a contiguous, array-backed representation with $O(1)$ access and specialized unboxed storage for primitives. In particular, **Int** values are stored directly in **Array**[**Int**], eliminating boxing, pointer indirection, and associated garbage-collection costs. This optimisation improves both cache efficiency and overall performance and is applied uniformly across all matchers, including **BruteForceMatcher** and **StatefulTreeMatcher**.

**Efficient iteration** All matching algorithms have performance-critical loops. The original JoinActors in [17] used Scala **for**-each loops, which are notoriously inefficient, as they are desugared into code that allocates closures with significant overhead [9]:

---

[6] Each **MatcherUtils**[**M**] instance also provides a *filtering predicate* derived from the analysis of the join pattern guard: it is used for the partial evaluation of guards discussed in Section 4.2.2.

▪ **Listing 9** Efficient **for**-each loops using opaque types and (inline) extension methods.

```
1  opaque type FastIterable[T] = IterableOnce[T]
2
3  extension[T] (it: IterableOnce[T])
4    def fast: FastIterable[T] = it
```

```
1  extension[T] (e: FastIterable[T])
2    inline def foreach[U](inline f: T => U): Unit =
3      val it = e.iterator
4      while it.hasNext do f(it.next())
```

```
val numbers = List(1, 2, 3)

for x <- numbers do
  f(x)
```
⇒ desugars to ⇒
```
val numbers = List(1, 2, 3)
val foreachFunc = new Function1[Int, Unit] {
  def apply(x: Int): Unit = f(x) }
numbers.foreach(foreachFunc)
```

We therefore wanted to replace such **for**-each loops with **while**-loops, which compile to efficient bytecode. To perform this replacement without sacrificing code clarity, we took advantage of several Scala 3 features: inspired by [3, 9], we introduced an *opaque type* **FastIterable**[T] (Listing 9) that wraps **IterableOnce**[T] without runtime overhead, providing an explicit opt-in conversion via the *extension method* fast. We then defined an *inline extension method* foreach that, when invoked on an expression of type **FastIterable**[T] by the desugaring of a **for**-each loop, expands into a **while**-loop:

```
val numbers = List(1, 2, 3)

for x <- numbers.fast do
  f(x)
```
⇒ expands to ⇒
```
val numbers = List(1, 2, 3)
val it = numbers.iterator
while it.hasNext do
  f(it.next())
```

**Lazy ramification**   As discussed in Section 2.2 and Figure 1, the **StatefulTreeMatcher** algorithm [17] maintains a tree-like state that remembers partial matches, and is updated ("ramified") as new messages arrive, while guaranteeing that a depth-first traversal of the tree always yields the fairest possible match (if any). The **Stateful-TreeMatcher** ramifies the entire matching tree for each new message, and then performs a depth-first traversal to identify complete matches, check their guards, and report the fairest match (if any). We implemented a different, *lazy* ramification strategy: when a new message arrives, we ramify the matching tree in depth-first order, and whenever the new message completes a partial match, we immediately evaluate its guard; if the result is **true**, we halt the ramification and report the successful match. This short-circuiting optimisation avoids the unnecessary growth of the match tree. We implemented it in the **WhileLazyMatcher**, **LazyParallelMatcher**, and **FilteringParallelMatcher**.

**Parallel ramification**   We parallelise the ramification of the matching tree by partitioning it into $n$ subtrees, each processed by a separate thread (where $n$ is the number of hardware threads supported by the system CPU). The partitions are constructed and ordered by fairness, i.e., for any two partitions $p$ and $p'$, if $p < p'$, then any match found in $p$ is fairer than any match found in $p'$. When a new message arrives, each thread performs a lazy ramification (described above) of its partition; whenever a thread reports a match, the threads which are ramifying less-fair partitions are interrupted. When all threads terminate, the fairest reported match (if any) is selected. This optimisation is implemented in the **LazyParallelMatcher** and **FilteringParallelMatcher**.

**Partial evaluation of guards**   The performance of join pattern matching can be improved by quickly recognising incoming messages that cannot possibly satisfy a guard. To achieve this, we identify *filtering clauses*: when a guard is a conjunction of predicates, we classify as filtering clauses those predicates that only depend on the payload of a single message type that only occurs once in the pattern. For example, consider the following join pattern:

```
1  Foo(x : Int) &:& Bar(y : Int) &:& Bar(z : Int)  if  x == 10 && y == 20 && z == x  =>  ...
```

Here, x == 10 is a filtering clause: it depends only on the payload of **Foo**(x), and such message type only occurs once in the pattern. Filtering clauses can be evaluated as soon as a corresponding message arrives — and if a message falsifies its filtering clause (e.g., the message **Foo**(42) falsifies x == 10), then the message can be ignored (without ramifying the matching tree), since it cannot be part of any message combination that could satisfy the guard. Note that the predicates y == 20 and z == x are *not* filtering clauses, because they depend on the payloads of **Bar**(y) and **Bar**(z) — i.e., they depend on a message type **Bar** that occurs multiple times in the pattern. As a consequence, if the message **Bar**(42) arrives, we do *not* discard it, even though it falsifies the predicate y == 20; in fact, that message could still be part of a message combination that may satisfy the predicate z == x (if x == 42), hence discarding **Bar**(42) would be incorrect. Compared to [17], we extend the receive macro to analyse the AST of the guards, extract their filtering clauses, and generate *filtering closures* that can be applied to incoming messages. Such filtering closures are used by the **FilteringParallelMatcher**, and this optimisation can reduce the size of the matching tree and its ramification costs.

## 5   Evaluation via Comparative Benchmarks

We now evaluate our improved **JoinActors** library by measuring the performance of its join pattern matching algorithms. Most of our evaluation is based on the benchmark suite originally presented in [17] for assessing the performance of join pattern matching. The rationale for this choice is that, to the best of our knowledge, this benchmark is the only one specifically designed to evaluate join pattern matching performance. In particular, the benchmark in [17] systematically varies the pattern size, guard complexity, and incoming message traffic (with varying amounts of out-of-order or unmatchable messages). Moreover, the benchmark also presents a worst case analysis showcasing that matching time can be exponential in the number of messages in the actor mailbox: this suggests that optimisations can be crucial for achieving reasonable performance. The evaluation includes synthetic benchmarks (Section 5.1), an adaptation of a smart house monitoring scenario (Section 5.2), and a bounded buffer benchmark (Section 5.3). For every benchmark, we measure the time needed to perform a fixed number of join pattern matches, and the throughput in number of matches per second; for comparison, the results include baselines measured using the version of **JoinActors** presented in [17]. Furthermore, in Section 5.4 we compare **JoinActors** to a popular actor library: Apache Pekko.

This evaluation also highlights the modularity of the JoinActors library design, which allows us to easily switch between different matching algorithms and systematically compare their performance. Moreover, the modular design is leveraged by an extensive test suite that, after fixing some join patterns and input traffic, switches between matching algorithms and checks that every one produces the same matches as the original implementation in [17]. This is also important for practical applications: our evaluation (like the one in in [17]) highlights that the efficiency of a join pattern matching algorithm is highly dependent on the specific workload and on the shape of the join patterns.

**Experimental setup**   We ran the experiments on a computer with Intel Core i7-1185G7 CPU (4 cores / 8 threads, 4.8 GHz) and 32 GB of RAM, with Pop!_OS 22.04. We used OpenJDK 25 with default settings, maximum heap size 16 GB, and Scala 3.7.3. Each benchmark is preceded by a warm-up phase that runs the same benchmark logic for 5 repetitions, to let the JVM perform optimisations such as JIT compilation and garbage collection tuning.

**5.1 Synthetic Benchmarks**

These benchmarks measure the performance of a single actor that attempts to match the messages it receives using exactly one join pattern. The messages are sent sequentially to the actor's mailbox, without delay. Timing starts with the sending of the first message; then, the actor performs 10 join pattern matches (which are guaranteed to occur by the content of the incoming traffic), and stops the timing. The size of the join pattern used by the actor (i.e., the number of messages complete the pattern) ranges from 1 to 5. We have two variations of the synthetic benchmarks, detailed below.

**Performance of join pattern matching without guards**   In this benchmark, the actor has a join pattern without a conditional guard. For instance, an actor with a guard-less join pattern of size 5 looks as follows:

```
1  Actor[…, …] {
2      receive { self =>
3          case A() &:& B() &:& C() &:& D() &:& E()   =>   …
4  }}}
```

The benchmark evaluates two workloads: one where the actor only receives "clean" messages that can be matched by the join pattern Figure 4, and another including additional "noise" messages that cannot be matched. Figure 5 presents the results. In general, our optimised matchers outperform the best performing baselines from [17] up to a factor of 3.8 for benchmarks with "clean" workload, and up to a factor of 31 for benchmarks with noise messages; for join patterns without guards, the "simpler" stateful matchers (i.e. **WhileLazyMatcher** and **StatefulTreeMatcher**) outperform the others. (For a detailed breakdown of the improvement over [17], see Table 2 and Table 3 in Appendix B.) Notably, on "clean" matchable messages, the baseline stateless **BruteForceMatcher** outperforms the baseline **StatefulTreeMatcher** from [17], due to the overhead of state maintenance — but our optimisations produce the opposite result, as they significantly reduce the cost of state maintenance. A similar phenomenon can be observed in the next benchmarks.

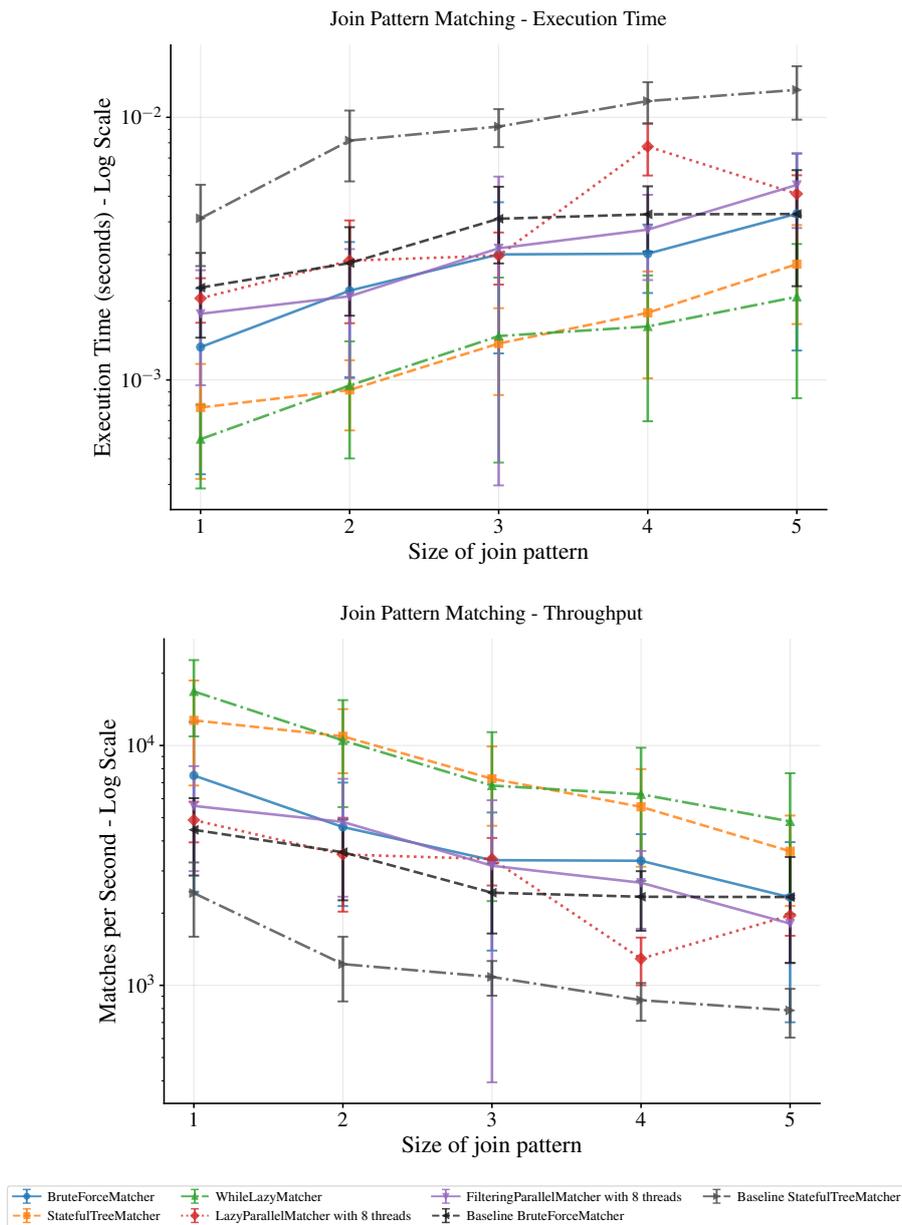

**Figure 4** Performance of join pattern matching without guards: with "clean" matchable messages. The plots show the average of 5 repetitions, with error bars indicating the standard deviation.

**Performance of join pattern matching with guards** In this benchmark, the actor has a join pattern which includes a conditional guard. For instance, an actor with a guarded join pattern of size 5 looks as follows:

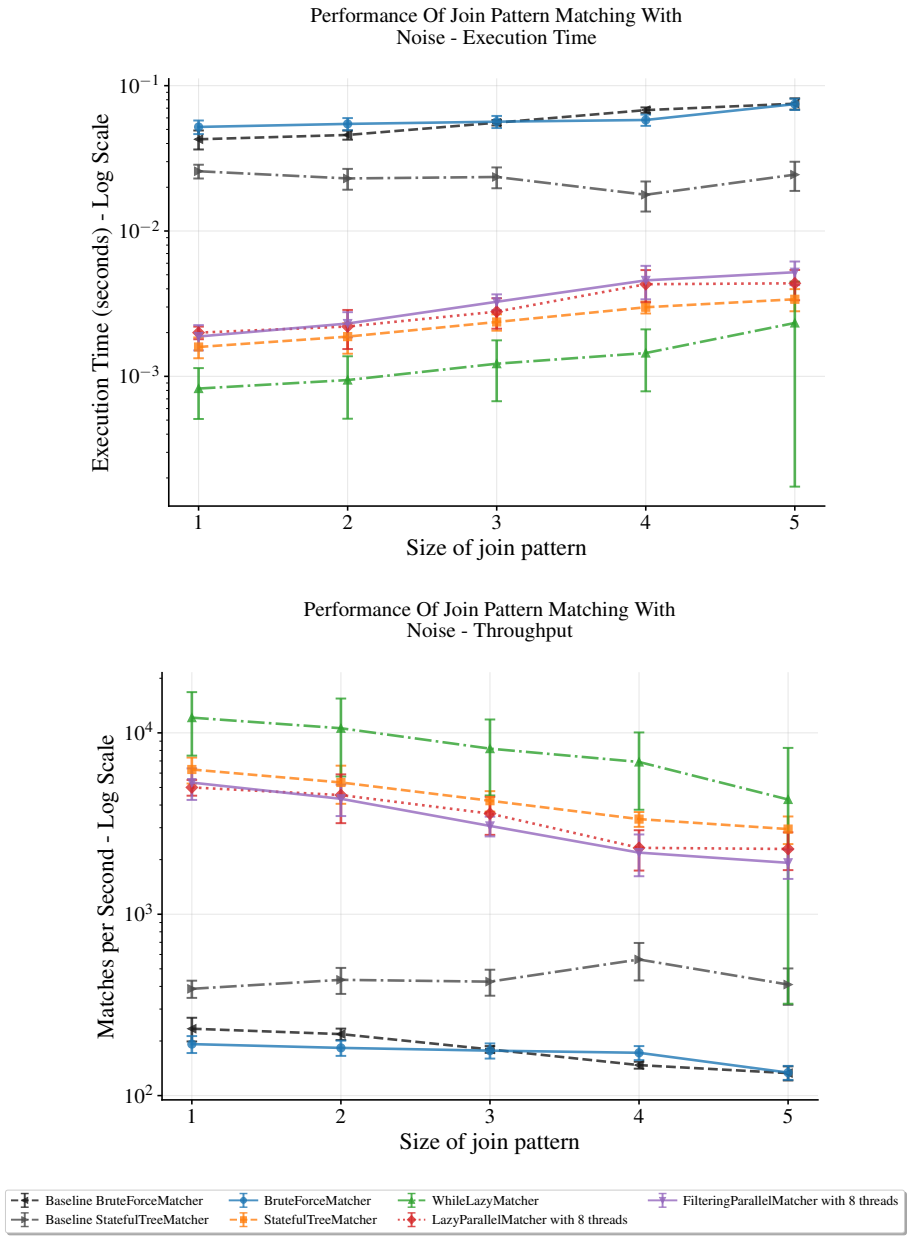

**Figure 5** Performance of join pattern matching without guards: with unmatchable "noise" messages. The plots show the average of 5 repetitions, with error bars indicating the standard deviation.

JoinActors: A Modular Library for Actors with Join Patterns```
1  Actor[…, …] {
2      receive { self => {
3          case A(x) &:& B(y) &:& C(z) &:& D(w) &:& E(a)   if   x == y && y == z && z == w && w == a   =>  …
4  }}}
```

The benchmark evaluates three workloads:

1. a stream of "clean" matchable messages;
2. a stream including additional "noise" messages of a type that cannot be matched (e.g., messages of type F for the actor above); and
3. a stream including "noise" messages whose type might be matched by the join pattern, but whose payloads do not satisfy the guard.

Figure 6, Figure 7, Figure 8 show the resulting measurements. Again, our optimised matchers generally outperform the best performing baselines from [17] up to a factor of 4.25 for the "clean" workload, up to a factor of 22 for the "noisy" workload, and up to a factor of 9 for the workload with non-matching payloads. (For more details, see Table 4, Table 5, and Table 6 in Appendix B.) When the "noise" consists of unmatchable message types, the simpler stateful matchers (i.e., WhileLazyMatcher and StatefulTreeMatcher) outperform the others — but when the "noise" consists of messages with non-matching payloads, the parallel matchers (i.e., LazyParallelMatcher and FilteringParallelMatcher) take the lead; note that FilteringParallelMatcher is slower because the guard does not contain any filtering clause.

### 5.2 Smart House Benchmark

This benchmark is adapted from a case study presented in [29] (and reused by [17]): an actor models a smart home monitor that receives messages from multiple household devices, and recognises certain combinations via non-trivial join patterns with guards. Notably, the join patterns in this benchmark overlap significantly, as many patterns share the same messages.[7] This overlap impacts matching performance, as a single message may create partial matches in multiple match trees (when using a stateful matcher), or may cause multiple patterns to be satisfied at the same time, thus requiring their comparison to find the fairest. (More details on this benchmark are available in Appendix A.)

The benchmark measures matching performance when the input messages contain varying amount of "noise", i.e., messages carrying random (and likely unmatchable) payloads. In Figure 9, the $x$-axis shows the number of "noise" messages sent alongside each group of 3 matchable messages. Again, our optimised matchers outperform the best baseline from [17] up to a factor of 58 as the number of "noise" messages increases. (See Table 7 in Appendix B for a detailed breakdown.) The FilteringParallelMatcher outperforms the other algorithms: this is because the patterns include filtering clauses that allow for quickly discarding "noise" messages, as shown in Listing 10.

---

[7] The bounded buffer benchmark in Section 5.3 also exhibits an overlap between join patterns, but such patterns are fewer and much simpler than the ones in the smart house benchmark.



A. Hussein, P. Haller, I. Karras, H. Melgratti, A. Scalas, E. Tuosto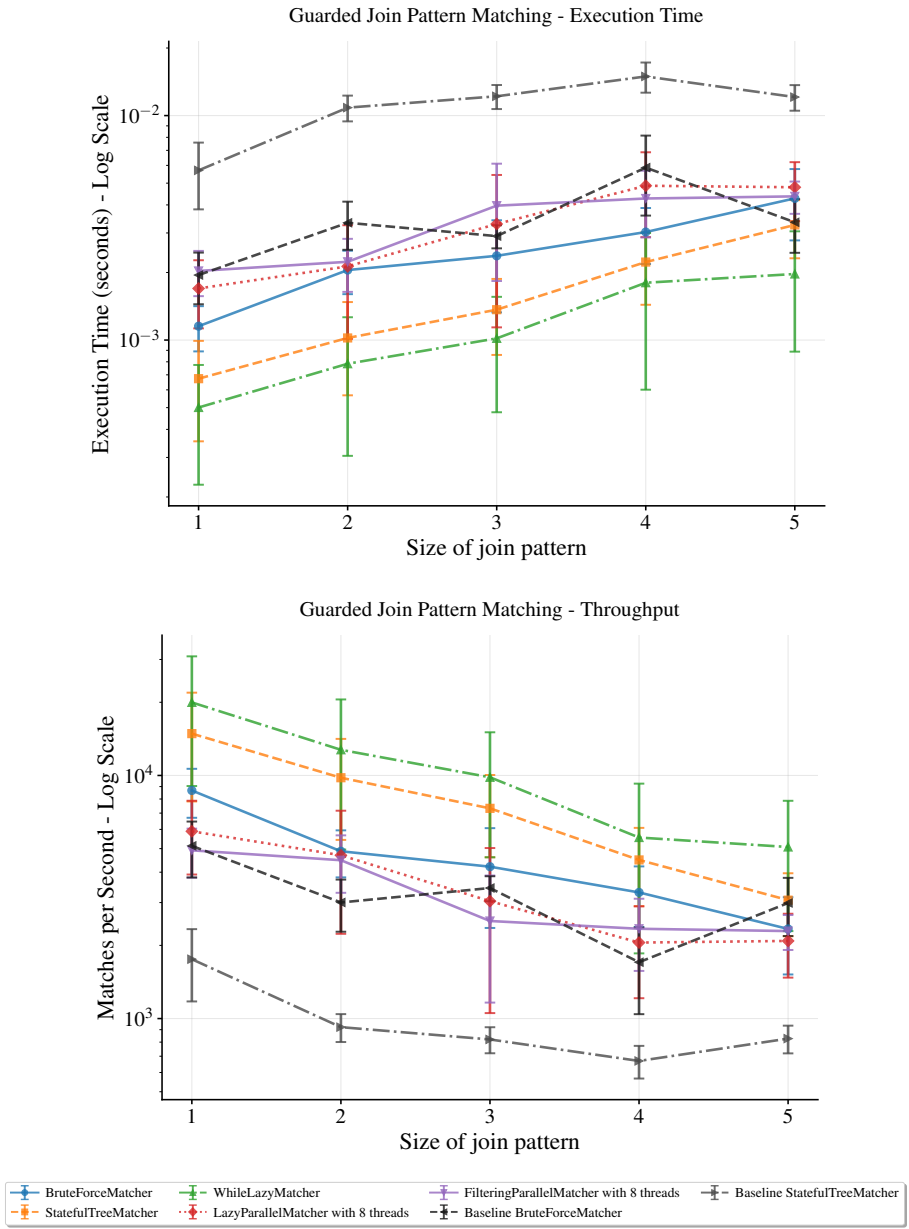

**Figure 6** Performance of join pattern matching with guards: with "clean" matchable messages. The plots show the average of 5 repetitions, with error bars indicating the standard deviation.

4:21



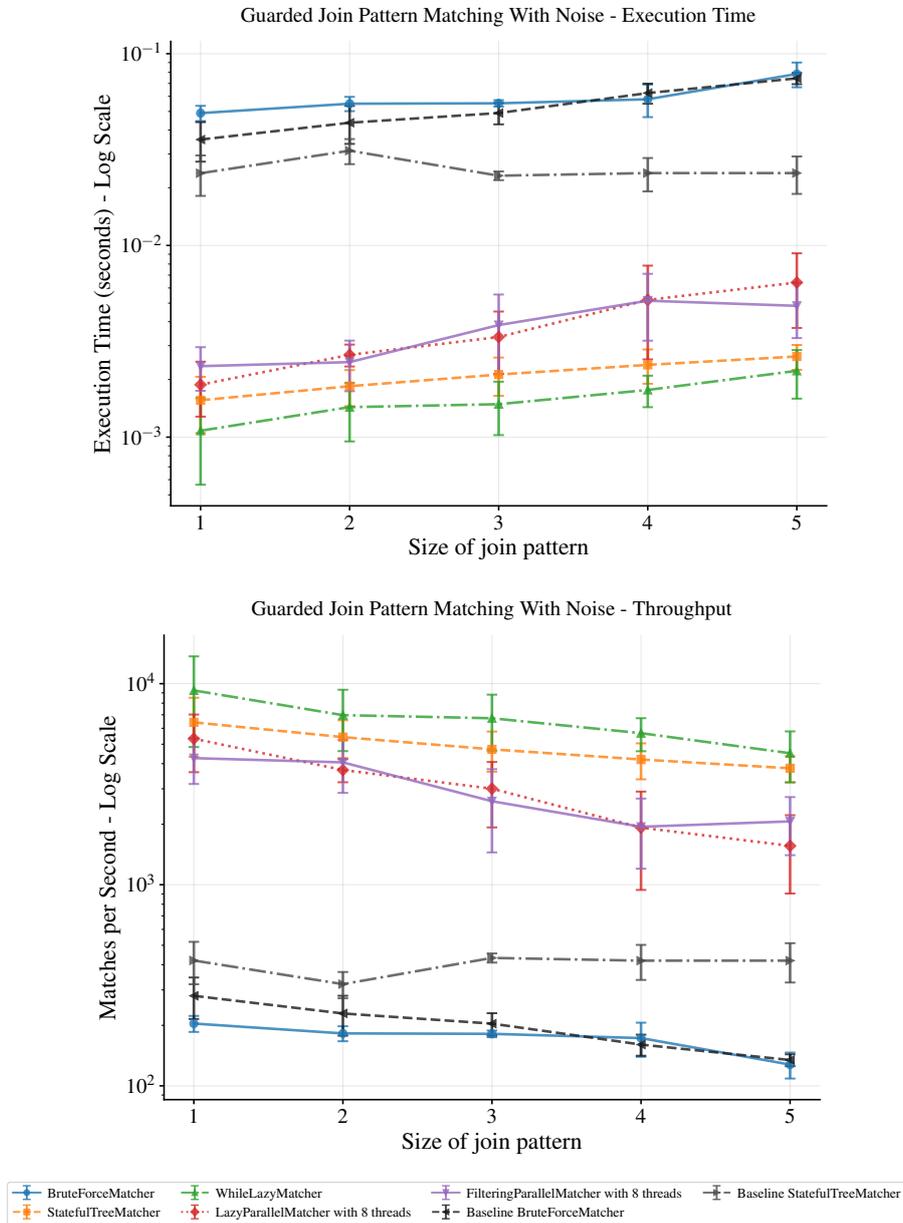

**Figure 7** Performance of join pattern matching with guards: with unmatchable "noise" messages. The plots show the average of 5 repetitions, with error bars indicating the standard deviation.





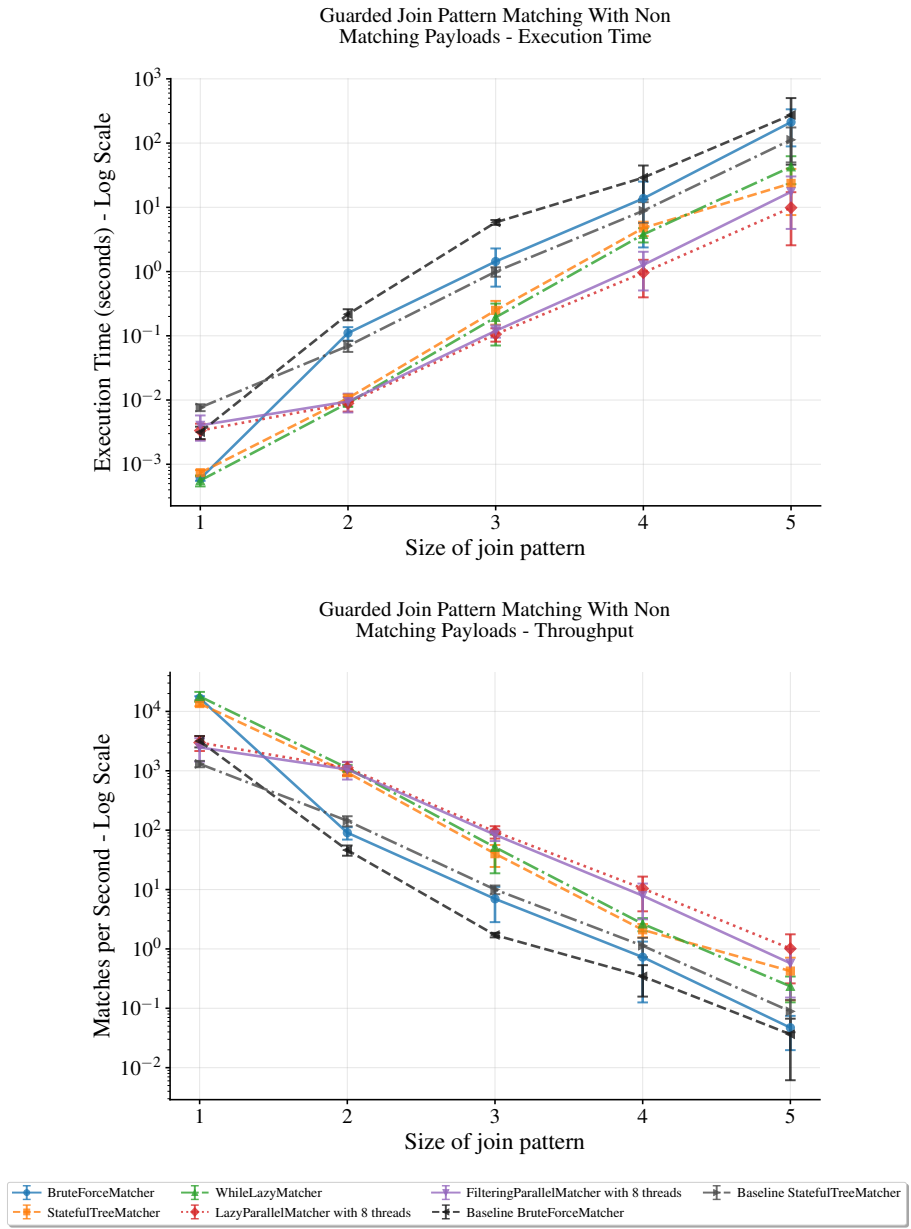

**Figure 8** Performance of join pattern matching with guards: with non-matching payloads. The plots show the average of 5 repetitions, with error bars indicating the standard deviation.





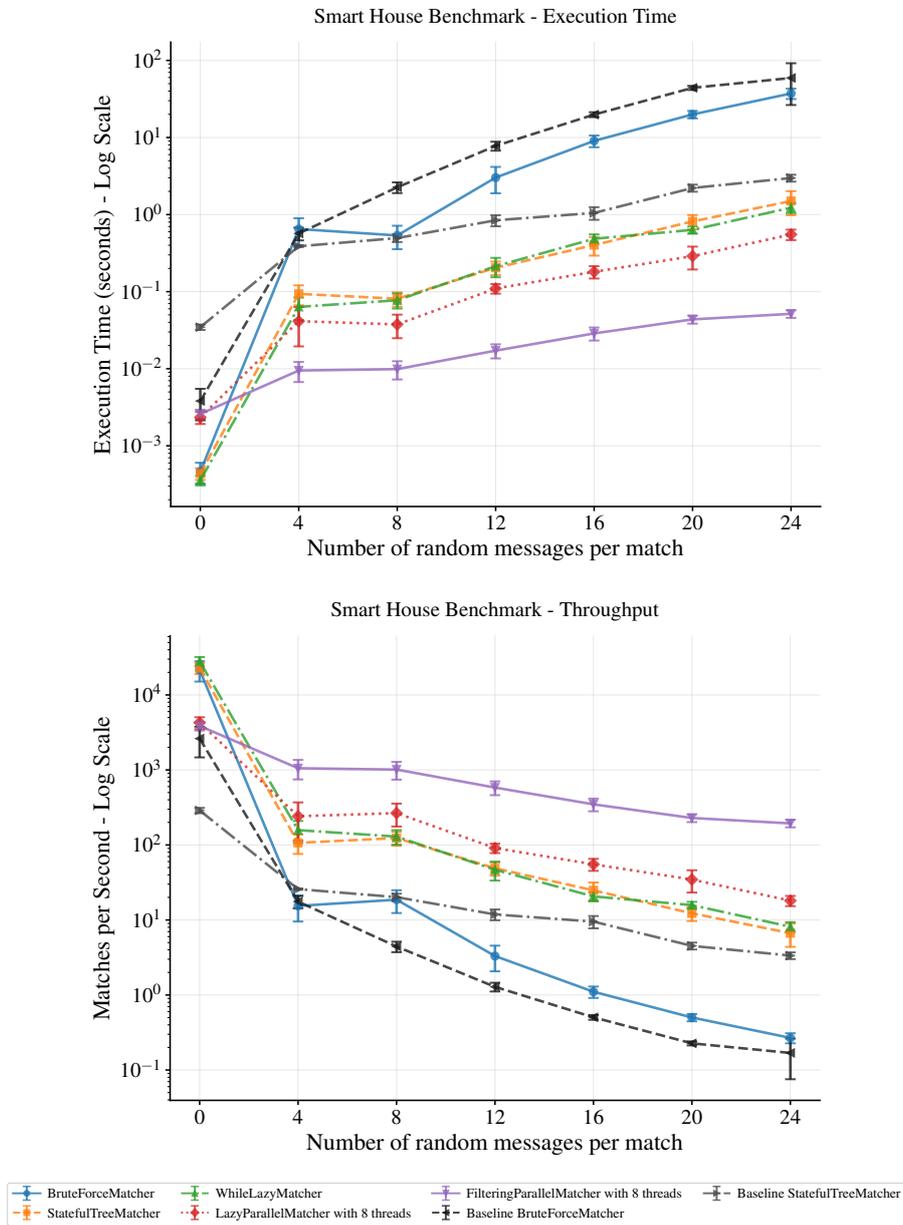

**Figure 9** Performance of the smart house benchmark. The plot show the average of 5 repetitions, with error bars indicating the standard deviation.



A. Hussein, P. Haller, I. Karras, H. Melgratti, A. Scalas, E. Tuosto■ **Listing 10** A simplified join pattern from the smart house monitor benchmark: it matches two **Motion** messages and a **Contact** message, and checks if the home is occupied based on their payloads, which include the rooms where the events were detected. Notably, the guard occupiedHome has the structure of a conjunction of predicates, and the one depending on cRoom0 is a filtering clause that can be leveraged by the **FilteringParallelMatcher** (Section 4.2.2). E.g., the matcher immediately discards a message like **Contact**(..., "gate_door", ...) because it would bind cRoom to "gate_door", which would falsify the filtering clause cRoom == "front_door".

```
1  inline def occupiedHome = (..., mRoom0: String, mRoom1: String, cRoom: String ) =>
2      ... && mRoom0 == "front_door" && cRoom == "front_door" && mRoom1 == "entrance_hall"
3
4  ...
5  case Motion(...., mRoom0: String, ...) &:& Contact(..., cRoom: String, ...) &:& Motion(..., mRoom1: String, ...)
6       if occupiedHome(..., mRoom0, mRoom1, cRoom)   =>   ...
```

### 5.3 Producers-Consumers Bounded Buffer Benchmark

This benchmark (originally part of the Savina suite [20] and later adapted in [17]) models the classical producer-consumer problem: producers and consumers are actors, and a buffer actor coordinates their interactions through join patterns: the buffer actor tracks whether the buffer is full or empty, queues consumers when data is unavailable, queues producers when the buffer is full, and notifies producers when buffer space becomes available. This benchmark evaluates the JoinActors performance under a high level of concurrency and contention.

Figure 10 presents the benchmark results for a buffer size of 1000 with 10 producers and 10 consumers. Also in this case, our optimised matchers outperform the baselines from [17] up to a factor of approx. 277, see Table 8 for a detailed breakdown. The **WhileLazyMatcher** achieves the best performance, which is expected, as this benchmark uses rather simple patterns without guards that do not leverage the optimisations of the parallel algorithms.

### 5.4 Evaluation of JoinActors as a Simple Actor Library

In this section, we compare the performance of JoinActors to two actor implementations that do *not* support join pattern matching. The aim is to evaluate the overhead introduced by JoinActors's join pattern matching mechanism, by applying it to cases where a simple pattern matching on each incoming message would be sufficient.

As a reference for our comparisons, we selected the popular actor library Apache Pekko[8] (which does not support join patterns). We adopted a third-party Pekko implementation of some of the benchmarks of the Savina actor benchmarking suite [20]:

---
[8] https://pekko.apache.org/ (Visited on 2026-02-12).

4:25



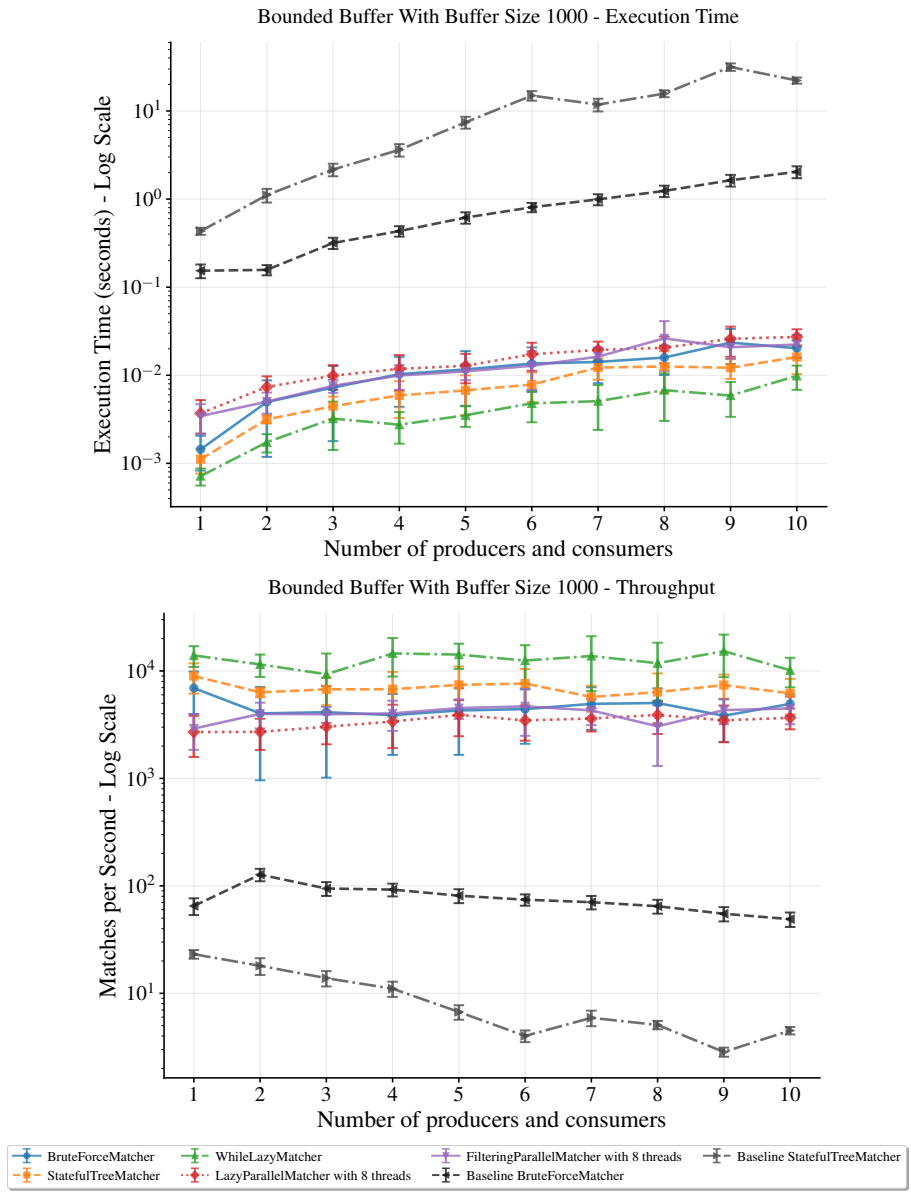

**Figure 10** Performance of the bounded buffer benchmark with buffer size 1000 and 10 producers and 10 consumers. The plot show the average of 5 repetitions, with error bars indicating the standard deviation.





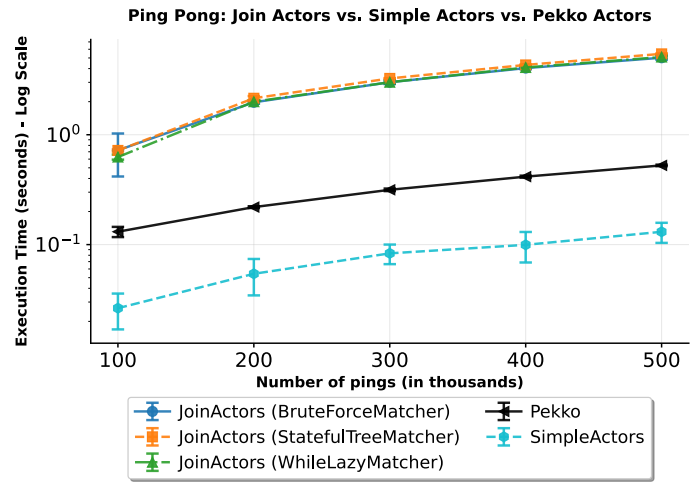

**Figure 11** Performance of a ping-pong actor benchmark using JoinActors join patterns. The plot show the average of 5 repetitions, with error bars indicating the standard deviation.

specifically, we took some implementations from the artifact of [27].[9] Then, we translated each adopted benchmark into the closest Actor equivalent under JoinActors, using different join pattern matching algorithms. Finally, we also implemented each selected benchmark using SimpleActor, which is a barebones actor implementation that uses the same minimal runtime of JoinActors, but only performs standard pattern matching on each incoming message, similarly to Pekko (see Appendix C). The rationale is that, when using simple pattern matching on individual incoming messages:

1. The Pekko implementations provides the reference performance for a widely-used, optimised actor library with a powerful runtime.
2. The SimpleActor-based implementations approximate the best-case performance achievable with a minimal runtime.
3. The JoinActors Actor-based implementations demonstrate the overhead introduced by the join pattern matching mechanism over SimpleActor.

The results of our comparison are as follows.

**Ping-Pong Benchmark (Figure 11)** Two actors send messages back and forth to each other, and the benchmark measures the time taken to complete a fixed number of message exchanges. This benchmark is highly dependent on the message matching performance, and shows the overhead introduced by the join pattern matching mechanics of JoinActors, which is slower than both Pekko and SimpleActor.

**Chameneos Benchmark (Figure 12)** Multiple chameneos (actors) interact with each other by changing colors upon meeting, and the benchmark measures the time taken

---

[9] Such third-party benchmark implementations are also available on https://github.com/dplyukhin/savina (Visited on 2026-02-12).





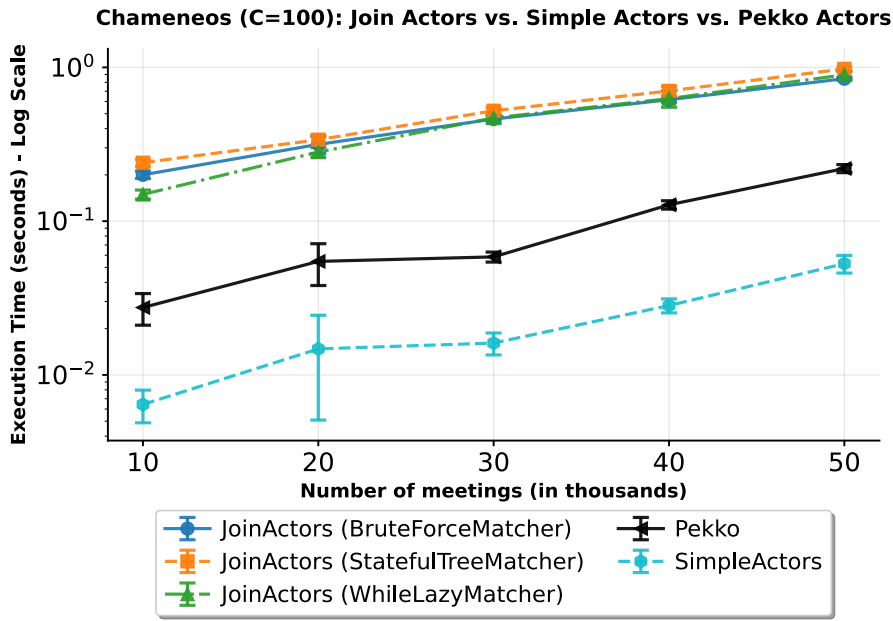

**Figure 12** Performance of a chameneos actor benchmark using JoinActors join patterns. The plot show the average of 5 repetitions, with error bars indicating the standard deviation.

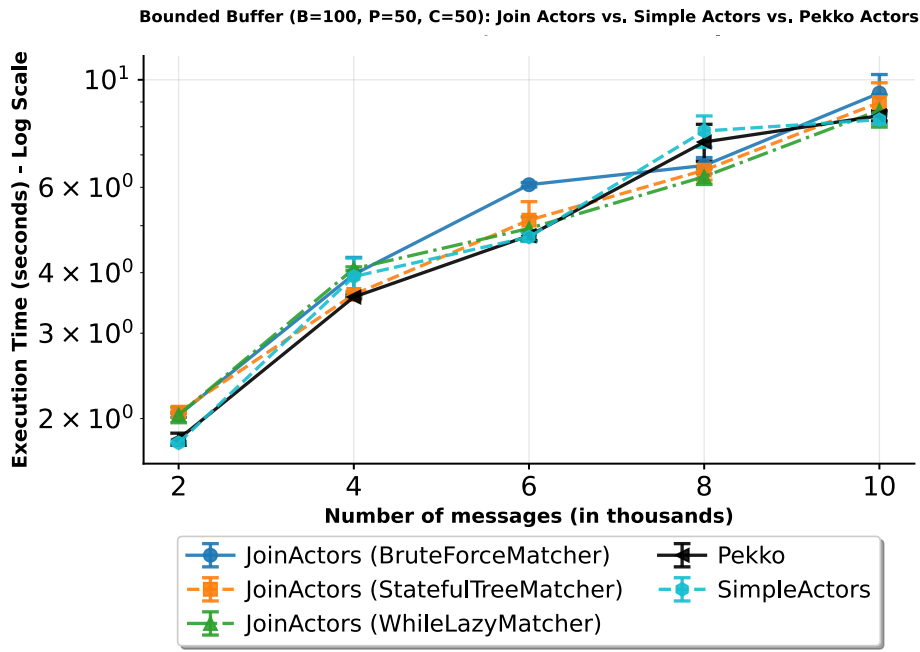

**Figure 13** Performance of a bounded buffer actor benchmark using JoinActors join patterns. The plot show the average of 5 repetitions, with error bars indicating the standard deviation.





to perform a fixed number of meetings. Similarly to ping-pong, this benchmark is highly dependent on the message matching performance.

**Bounded Buffer Benchmark (Figure 13)**    This benchmark implements a bounded buffer actor that coordinates multiple producer and consumer actors.[10] The benchmark measures the time taken to process a fixed number of messages between producers and consumers through the buffer; the buffer size is set to 100, with 50 producers and 50 consumers. This benchmark is influenced by the thread contention between producer and consumer actors, so the raw message matching performance is less impactful than the other benchmarks. As a consequence, the plots show similar performance for Pekko and JoinActors's **Actor**s and **SimpleActor**.

In conclusion, our measurements highlight that, in the presence of actors that perform simple one-message-at-a-time matching:

1. The join pattern matching mechanism of JoinActors introduces overhead, especially visible in benchmarks heavily reliant on matching speed (Figure 11 and Figure 12).[11]
2. If factors other than message matching performance become relevant, such as thread contention, then the overhead of JoinActors's join pattern matching mechanism is less impactful (Figure 13).
3. When matching one message at a time, the choice of join pattern matching algorithm becomes irrelevant, as none of our optimised algorithms performs better than the **BruteForceMatcher**. (For clarity, the plots only show part of the matching algorithms available in JoinActors, but the remaining ones have similar or worse performance.)

## 6   Related Work

Since its introduction, the join calculus [13] has inspired a wide variety of implementations across programming languages [2, 7, 10, 12, 14, 15, 16, 21, 23, 25, 26, 29, 30, 31, 35, 36]. The first implementations closely adhered to the original join calculus and did not support pattern matching over message payloads; then, [24] extended join patterns with constant-value matching, and [23] generalised this idea to algebraic data structures. More expressive forms of pattern matching were later studied in [16, 26, 29]. Unlike [23], these systems support join patterns with *guards* resembling those in functional programming languages such as Erlang, Scala, or F#. The introduction of guards improves the expressiveness and usability of join patterns, but also increases implementation complexity. To support guards, [16] adopts a stateless brute-force

---

[10] Note that this implementation of bounded buffer is different from the one in Section 5.3: the latter is designed to use join patterns to match combinations of multiple producer/consumer messages, whereas the Savina version only uses simple matching of one message at a time.
[11] In principle, our receive macro (Section 4.2.1) could identify whether a given set of join patterns only matches one message at a time, and thus, could leave that code unchanged, thus achieving a performance similar to **SimpleActor**. However, we have not yet implemented this optimisation.





strategy, while [26] maintains caches of partial matches to reduce redundant computation. Instead, [29] employs a variant of the RETE algorithm [11], maintaining a *discrimination network* of partial matches — although this implementation is not publicly available. Later, [17] introduced new matching algorithms (stateful and stateless) that implement a *deterministic and fair* matching policy.

The closest related work to this paper are the language-level embeddings of the join calculus and join pattern matching:

1. JoCaml [12] extends OCaml with join patterns and message-passing concurrency primitives. Its runtime employs threads, with synchronisation achieved via message-passing rather than shared memory.

2. JErlang [26] extends Erlang with join-based synchronisation, addressing the language's limited native support for multi-process coordination. JErlang is implemented as a library that requires a modified Erlang VM that incorporates RETE-based optimisations to improve efficiency.

3. Sparrow [29] is a domain-specific language embedded in Elixir via macros. It extends the actor model from single-message to multi-message matching and draws on techniques from complex event processing frameworks. It uses an adapted version of the RETE algorithm for join pattern matching.

4. Scala Joins [16] provide a library-based implementation for Scala 2, leveraging extensible pattern matching, and primarily targeted toward actor-based applications.

In the join calculus, join patterns are interpreted under a nondeterministic matching policy: when multiple message combinations are enabled, one is selected nondeterministically. The implementations listed above (except [17]) either leave the matching policy unspecified or adopt a first-match strategy [35]. The issue is partially addressed in [5], which introduces *parallel object monitors* (POMs) for coordinating concurrent objects via *schedulers* that implement specific message selection policies. In particular, [5] discusses possible applications of POMs to tame the nondeterministm of join pattern matching: it sketches a bounded buffer implementation in a variation of Polyphonic C# [2] (which is an exension of C# with constructs inspired by join patterns) using a POM scheduler to ensure that read/write requests are selected in order of arrival, rather than nondeterministically. Instead, [17] introduces and formalises the fair, deterministic join pattern matching policy implemented in the first proof-of-concept version of JoinActors. In this work, we revisit and improve JoinActors and present a comprehensive description of its modular and extensible design; we extend it with additional join pattern matching algorithms (still based on the fair matching policy in [17]) showing significantly improved performance.

## 7 Conclusion and Future Work

We presented JoinActors, a modular and extensible actor library that brings guarded join patterns to Scala 3. We showcased its advanced use of Scala 3's metaprogramming facilities, which allows JoinActors to implement guarded join patterns as regular Scala code, without requiring any modifications to the compiler or the language itself.





Leveraging JoinActors's modularity, we also presented new optimised join pattern matching algorithms with significant performance improvements over the early JoinActors prototype introduced in [17]. Our work demonstrates the usability of JoinActors as a research playground for further exploration of join pattern matching algorithms and experimentation with new features.

**Future work** Building upon the modular and extensible architecture of JoinActors, we identify several promising directions for further research and development.

First, we plan to study further optimisations. For instance, the receive macro could be optimised to detect when join patterns only match one message at a time (i.e., when they behave like simple pattern matching on individual messages) and bypass the join pattern matching machinery in such cases, making the performance of JoinActors's Actor comparable to that of SimpleActor and Apache Pekko (Section 5.4). Moreover, parallelisation has demonstrated measurable performance improvements on selected benchmarks: see Figure 9, where the LazyParallelMatcher and FilteringParallelMatcher achieved the best results. This outcome indicates that parallelisation alone, even without the guard filtering of FilteringParallelMatcher, can provide performance gains compared to sequential algorithms. However, the other benchmarks suggest that the effectiveness of parallelisation is highly workload-dependent and not universally applicable. An in-depth performance analysis of our current matching algorithms (e.g. using profiling tools) could provide further insights into their behaviour, guiding both implementation choices and theoretical understanding – and indeed, the optimisations presented in this work were selected by iteratively using profiling data to identify the most promising directions for improving performance [22]. An ablation study (with the systematic addition or removal of individual optimisations, leveraging JoinActors's modular design) could further pinpoint which optimisation contributes most to the join pattern matching performance, depending on the workload.

Another source of inspiration for extensions and optimisations to JoinActors is the literature on complex event processing (CEP) [4, 6, 28]. On the one hand, some techniques from CEP could be used to optimise join pattern matching further (although such optimisations would likely need to be adapted to support our fair matching requirement). On the other hand, JoinActors could provide a basis to perform CEP with dynamically reconfigurable rules: this is because the join pattern matching structures generated by JoinActors's receive macro (via generateJoinPattern, explained in Section 4.2.1) could be also instantiated at runtime (although JoinActors does not yet provide a documented API to achieve this).

Future versions of JoinActors could explore enhancements such as recursive pattern matching over algebraic data types, bound actor mailboxes to improve resource control, and more sophisticated partial evaluation of complex guards.

Thanks to the modular design of JoinActors, most of these research directions can be pursued by implementing new join pattern matching algorithms that can exist alongside the existing ones, as outlined in Section 4.1.

We also plan to investigate how the design of JoinActors could be ported to other languages. A natural candidate is Rust, given its focus on safety and performance, its advanced type system, and its support for metaprogramming via macros.



**JoinActors: A Modular Library for Actors with Join Patterns**

**Data-Availability Statement**   The evaluated artifact for this paper contains all significant code examples and scripts for reproducing benchmark results. You can find the companion artifact in [19].

**Acknowledgements**   This work was inspired by the discussion on *"Join patterns/ synchronisation: the next generation"* [8, p. 54] at the Dagstuhl Seminar 21372. We thank Hadi Abdalla, Arooj Chaudhry, Bence Gattyan, Mohamed Ali Resho, Bozhi Lyu, Leonard Theisler, and Andrei Popa for co-designing the original payment system used as a reference for Listing 1. Research partly supported by: Horizon Europe grant 101093006 (TaRDIS), PRIN PNRR project DeLICE (P20223T2MF), MUR (Italy) Department of Excellence 2023-2027 for GSSI.





# A Smart House Monitor Example

The join patterns of the smart house monitoring example (Listing 11) exhibit the following behaviours:

**Lighting control:** whenever messages signaling motion detection and low ambient illuminance are received, turn the lights on;

**Presence detection:** recognise message sequences that signify residents' arrivals or departures and reacts accordingly.





■ **Listing 11** A smart house monitor and controller using the JoinActors library. The actor manages household devices by activating lights and detecting resident arrivals or departures.

```
 1  def smartHouseExample(matcherAlgorithm: MatcherFactory) =
 2    var lastNotification = Date(0L)
 3    var lastMotionInBathroom = Date(0L)
 4    def isSorted: Seq[Date] => Boolean = times =>
 5      times.sliding(2).forall { case Seq(x, y) => x.before(y) || x == y }
 6    inline def bathroomOccupied =
 7      ( times: Seq[Date], rooms: Seq[String], mStatus: Boolean, lStatus: Boolean, value: Int) =>
 8        isSorted(times) && rooms.forall(_ == "bathroom") && mStatus && !lStatus && value <= 40
 9
10    inline def occupiedHome =
11    (times: Seq[Date], statuses: Seq[Boolean], mRoom0: String, mRoom1: String, cRoom: String ) =>
12      isSorted(times) && statuses.forall(
13        _ == true
14      ) && mRoom0 == "front_door" && cRoom == "front_door" && mRoom1 == "entrance_hall"
15
16    inline def emptyHome = (times: Seq[Date], statuses: Seq[Boolean], mRoom0: String, mRoom1: String, cRoom: String)
        ↪    =>
17      isSorted(times) && statuses.forall( _ == true) && mRoom0 == "entrance_hall" && cRoom == "front_door" &&
        ↪    mRoom1 == "front_door"
18
19    Actor[Action, Unit] {
20      receive { (self: ActorRef[Action]) =>
21      { // E1. Turn on the lights of the bathroom if someone enters in it, and its ambient light is less than 40 lux.
22        case Motion(_: Int, mStatus: Boolean, mRoom: String, t0: Date)
23          &:& AmbientLight(_: Int, value: Int, alRoom: String, t1: Date)
24          &:& Light(_: Int, lStatus: Boolean, lRoom: String, t2: Date)
25            if bathroomOccupied(List(t0, t1, t2), List(mRoom, lRoom, alRoom), mStatus, lStatus, value ) =>
26          lastNotification = Date()
27          lastMotionInBathroom = lastNotification
28          println("Someone entered the bathroom")
29          Continue
30        // E5. Detect home arrival or leaving based on a particular sequence of messages, and activate the corresponding
        ↪    scene.
31        case Motion(_: Int, mStatus0: Boolean, mRoom0: String, t0: Date)
32          &:& Contact(_: Int, cStatus: Boolean, cRoom: String, t1: Date)
33          &:& Motion(_: Int, mStatus1: Boolean, mRoom1: String, t2: Date)
34            if occupiedHome(List(t0, t1, t2), List(mStatus0, mStatus1, cStatus), mRoom0, mRoom1, cRoom ) =>
35          lastNotification = Date()
36          println("Someone arrived home")
37          Continue
38        case Motion(_: Int, mStatus0: Boolean, mRoom0: String, t0: Date)
39          &:& Contact(_: Int, cStatus: Boolean, cRoom: String, t1: Date)
40          &:& Motion(_: Int, mStatus1: Boolean, mRoom1: String, t2: Date)
41            if emptyHome( List(t0, t1, t2), List(mStatus0, mStatus1, cStatus), mRoom0, mRoom1, cRoom) =>
42          lastNotification = Date()
43          println("Someone left home")
44          Continue
45
46        case ShutOff() =>
47          println("Shutting down the smart house. Bye!")
48          Stop(())
49      }
50    }(matcherAlgorithm)
51  }
```





## B Performance Improvement Compared to Baselines fron ECOOP'24

See Table 2, Table 3, Table 4, Table 5, Table 6, Table 7, and Table 8.

▰ **Table 2** Performance comparison for join pattern matching without guards.

| Algorithm | vs Baseline | \multicolumn{5}{c}{Join Pattern Size} | | | | |
|---|---|---|---|---|---|---|
| | | 1 | 2 | 3 | 4 | 5 |
| \multicolumn{7}{c}{**Baseline Performance (messages per second)**} | | | | | | |
| Baseline BruteForceMatcher | (reference) | 4451 mps | 3589 mps | 2434 mps | 2341 mps | 2335 mps |
| Baseline StatefulTreeMatcher | (reference) | 2424 mps | 1226 mps | 1084 mps | 867 mps | 786 mps |
| \multicolumn{7}{c}{**Improvement Factors**} | | | | | | |
| BruteForceMatcher | vs Baseline BruteForceMatcher | 1.68x | 1.27x | 1.37x | 1.41x | 1.00x |
| | vs Baseline StatefulTreeMatcher | 3.09x | 3.73x | 3.07x | 3.81x | 2.96x |
| StatefulTreeMatcher | vs Baseline BruteForceMatcher | 2.86x | 3.04x | 2.99x | 2.37x | 1.55x |
| | vs Baseline StatefulTreeMatcher | 5.25x | 8.91x | 6.70x | 6.40x | 4.61x |
| WhileLazyMatcher | vs Baseline BruteForceMatcher | 3.78x | 2.93x | 2.80x | 2.67x | 2.07x |
| | vs Baseline StatefulTreeMatcher | 6.93x | 8.56x | 6.28x | 7.22x | 6.14x |
| FilteringParallelMatcher | vs Baseline BruteForceMatcher | 1.26x | 1.34x | 1.30x | 1.14x | 0.77x |
| | vs Baseline StatefulTreeMatcher | 2.31x | 3.92x | 2.91x | 3.09x | 2.30x |
| LazyParallelMatcher | vs Baseline BruteForceMatcher | 1.10x | 0.98x | 1.38x | 0.55x | 0.84x |
| | vs Baseline StatefulTreeMatcher | 2.02x | 2.87x | 3.10x | 1.49x | 2.49x |





**Table 3** Performance comparison for join pattern matching without guards with noise

| Algorithm | vs Baseline | Join Pattern Size | | | | |
|---|---|---|---|---|---|---|
| | | 1 | 2 | 3 | 4 | 5 |
| **Baseline Performance (messages per second)** | | | | | | |
| Baseline BruteForceMatcher | (reference) | 234 mps | 218 mps | 180 mps | 147 mps | 133 mps |
| Baseline StatefulTreeMatcher | (reference) | 388 mps | 435 mps | 425 mps | 563 mps | 410 mps |
| **Improvement Factors** | | | | | | |
| BruteForceMatcher | vs Baseline BruteForceMatcher | 0.82x | 0.84x | 0.98x | 1.17x | 1.01x |
| | vs Baseline StatefulTreeMatcher | 0.50x | 0.42x | 0.42x | 0.31x | 0.33x |
| StatefulTreeMatcher | vs Baseline BruteForceMatcher | 26.87x | 24.41x | 23.51x | 22.72x | 22.16x |
| | vs Baseline StatefulTreeMatcher | 16.19x | 12.25x | 9.95x | 5.94x | 7.20x |
| WhileLazyMatcher | vs Baseline BruteForceMatcher | 51.90x | 48.58x | 45.54x | 46.93x | 32.26x |
| | vs Baseline StatefulTreeMatcher | 31.27x | 24.39x | 19.26x | 12.28x | 10.48x |
| FilteringParallelMatcher | vs Baseline BruteForceMatcher | 22.76x | 19.84x | 17.07x | 14.85x | 14.44x |
| | vs Baseline StatefulTreeMatcher | 13.71x | 9.96x | 7.22x | 3.88x | 4.69x |
| LazyParallelMatcher | vs Baseline BruteForceMatcher | 21.40x | 20.80x | 19.96x | 15.77x | 17.21x |
| | vs Baseline StatefulTreeMatcher | 12.89x | 10.44x | 8.44x | 4.13x | 5.59x |





▰ **Table 4** Performance improvement factors for guarded join pattern matching.

| Algorithm | vs Baseline | Join Pattern Size | | | | |
|---|---|---|---|---|---|---|
| | | 1 | 2 | 3 | 4 | 5 |
| **Baseline Performance (messages per second)** | | | | | | |
| Baseline BruteForceMatcher | (reference) | 5133 mps | 3003 mps | 3443 mps | 1705 mps | 2988 mps |
| Baseline StatefulTreeMatcher | (reference) | 1754 mps | 922 mps | 821 mps | 669 mps | 827 mps |
| **Improvement Factors** | | | | | | |
| BruteForceMatcher | vs Baseline BruteForceMatcher | 1.69x | 1.62x | 1.22x | 1.94x | 0.78x |
| | vs Baseline StatefulTreeMatcher | 4.94x | 5.29x | 5.14x | 4.93x | 2.83x |
| StatefulTreeMatcher | vs Baseline BruteForceMatcher | 2.90x | 3.26x | 2.13x | 2.63x | 1.03x |
| | vs Baseline StatefulTreeMatcher | 8.47x | 10.62x | 8.93x | 6.71x | 3.71x |
| WhileLazyMatcher | vs Baseline BruteForceMatcher | 3.89x | 4.25x | 2.85x | 3.26x | 1.70x |
| | vs Baseline StatefulTreeMatcher | 11.38x | 13.83x | 11.98x | 8.30x | 6.14x |
| FilteringParallelMatcher | vs Baseline BruteForceMatcher | 0.96x | 1.49x | 0.73x | 1.37x | 0.77x |
| | vs Baseline StatefulTreeMatcher | 2.81x | 4.86x | 3.07x | 3.50x | 2.77x |
| LazyParallelMatcher | vs Baseline BruteForceMatcher | 1.15x | 1.56x | 0.88x | 1.20x | 0.70x |
| | vs Baseline StatefulTreeMatcher | 3.36x | 5.09x | 3.71x | 3.07x | 2.52x |





▪ **Table 5** Performance improvement factors for guarded join pattern matching with noise messages.

| Algorithm | vs Baseline | Join Pattern Size | | | | |
|---|---|---|---|---|---|---|
| | | 1 | 2 | 3 | 4 | 5 |
| | **Baseline Performance (messages per second)** | | | | | |
| Baseline BruteForceMatcher | (reference) | 280 mps | 229 mps | 204 mps | 160 mps | 134 mps |
| Baseline StatefulTreeMatcher | (reference) | 420 mps | 320 mps | 433 mps | 419 mps | 419 mps |
| | **Improvement Factors** | | | | | |
| BruteForceMatcher | vs Baseline BruteForceMatcher | 0.73x | 0.80x | 0.89x | 1.08x | 0.95x |
| | vs Baseline StatefulTreeMatcher | 0.49x | 0.57x | 0.42x | 0.41x | 0.30x |
| StatefulTreeMatcher | vs Baseline BruteForceMatcher | 22.90x | 23.67x | 23.12x | 26.16x | 28.24x |
| | vs Baseline StatefulTreeMatcher | 15.28x | 16.92x | 10.89x | 10.01x | 9.05x |
| WhileLazyMatcher | vs Baseline BruteForceMatcher | 33.01x | 30.44x | 33.01x | 35.37x | 33.56x |
| | vs Baseline StatefulTreeMatcher | 22.03x | 21.76x | 15.55x | 13.53x | 10.75x |
| FilteringParallelMatcher | vs Baseline BruteForceMatcher | 15.20x | 17.72x | 12.76x | 12.11x | 15.37x |
| | vs Baseline StatefulTreeMatcher | 10.15x | 12.67x | 6.01x | 4.63x | 4.93x |
| LazyParallelMatcher | vs Baseline BruteForceMatcher | 18.99x | 16.24x | 14.73x | 12.00x | 11.62x |
| | vs Baseline StatefulTreeMatcher | 12.68x | 11.61x | 6.94x | 4.59x | 3.72x |







**Table 6** Performance improvement factors for guarded join pattern matching with non-matching payloads.

| Algorithm | vs Baseline | Join Pattern Size | | | | |
|---|---|---|---|---|---|---|
| | | 1 | 2 | 3 | 4 | 5 |
| | **Baseline Performance (messages per second)** | | | | | |
| Baseline BruteForceMatcher | (reference) | 3167 mps | 46 mps | 2 mps | 0 mps | 0 mps |
| Baseline StatefulTreeMatcher | (reference) | 1306 mps | 144 mps | 10 mps | 1 mps | 0 mps |
| | **Improvement Factors** | | | | | |
| BruteForceMatcher | vs Baseline BruteForceMatcher | 5.28x | 1.96x | 4.06x | 2.11x | 1.29x |
| | vs Baseline StatefulTreeMatcher | 12.81x | 0.63x | 0.70x | 0.64x | 0.53x |
| StatefulTreeMatcher | vs Baseline BruteForceMatcher | 4.28x | 20.26x | 23.44x | 6.08x | 11.61x |
| | vs Baseline StatefulTreeMatcher | 10.37x | 6.51x | 4.01x | 1.85x | 4.77x |
| WhileLazyMatcher | vs Baseline BruteForceMatcher | 5.65x | 24.11x | 30.10x | 7.66x | 6.42x |
| | vs Baseline StatefulTreeMatcher | 13.69x | 7.75x | 5.15x | 2.34x | 2.64x |
| FilteringParallelMatcher | vs Baseline BruteForceMatcher | 0.78x | 22.84x | 48.84x | 22.84x | 15.72x |
| | vs Baseline StatefulTreeMatcher | 1.90x | 7.34x | 8.35x | 6.97x | 6.46x |
| LazyParallelMatcher | vs Baseline BruteForceMatcher | 0.94x | 24.43x | 55.08x | 30.18x | 27.75x |
| | vs Baseline StatefulTreeMatcher | 2.29x | 7.85x | 9.42x | 9.21x | 11.40x |



■ **Table 7** Performance improvement factors for smart house benchmark.

| Algorithm | vs Baseline | Number of Random Messages per Match | | | | | | |
|---|---|---|---|---|---|---|---|---|
| | | 0 | 4 | 8 | 12 | 16 | 20 | 24 |
| **Baseline Performance (messages per second)** | | | | | | | | |
| Baseline BruteForceMatcher | (reference) | 2620 mps | 18 mps | 4 mps | 1 mps | 1 mps | 0 mps | 0 mps |
| Baseline StatefulTreeMatcher | (reference) | 289 mps | 26 mps | 20 mps | 12 mps | 10 mps | 5 mps | 3 mps |
| **Improvement Factors** | | | | | | | | |
| BruteForceMatcher | vs Baseline BruteForceMatcher | 8.24x | 0.88x | 4.21x | 2.57x | 2.19x | 2.21x | 1.59x |
| | vs Baseline StatefulTreeMatcher | 74.80x | 0.60x | 0.93x | 0.28x | 0.12x | 0.11x | 0.08x |
| StatefulTreeMatcher | vs Baseline BruteForceMatcher | 8.75x | 6.07x | 27.92x | 38.06x | 49.26x | 54.13x | 39.49x |
| | vs Baseline StatefulTreeMatcher | 79.40x | 4.14x | 6.13x | 4.12x | 2.61x | 2.72x | 1.99x |
| WhileLazyMatcher | vs Baseline BruteForceMatcher | 10.69x | 8.97x | 29.21x | 36.33x | 40.86x | 69.60x | 48.17x |
| | vs Baseline StatefulTreeMatcher | 97.00x | 6.11x | 6.41x | 3.94x | 2.17x | 3.49x | 2.43x |
| FilteringParallelMatcher | vs Baseline BruteForceMatcher | 1.47x | 59.98x | 228.52x | 453.02x | 690.27x | 1009.59x | 1148.28x |
| | vs Baseline StatefulTreeMatcher | 13.38x | 40.88x | 50.18x | 49.10x | 36.58x | 50.68x | 57.83x |
| LazyParallelMatcher | vs Baseline BruteForceMatcher | 1.63x | 13.74x | 60.07x | 70.79x | 109.53x | 152.61x | 107.18x |
| | vs Baseline StatefulTreeMatcher | 14.79x | 9.37x | 13.19x | 7.67x | 5.81x | 7.66x | 5.40x |







**Table 8** Performance improvement factors for bounded buffer benchmark.

| Algorithm | vs Baseline | | 1 | 2 | 3 | 4 | 5 | 6 | 7 | 8 | 9 | 10 |
|---|---|---|---|---|---|---|---|---|---|---|---|---|
| | | | Baseline Performance (messages per second) | | | | | | | | | |
| Baseline BruteForceMatcher | (reference) | | 65 mps | 127 mps | 94 mps | 92 mps | 81 mps | 74 mps | 71 mps | 65 mps | 55 mps | 49 mps |
| Baseline StatefulTreeMatcher | (reference) | | 23 mps | 18 mps | 14 mps | 11 mps | 7 mps | 4 mps | 6 mps | 5 mps | 3 mps | 5 mps |
| | | | Improvement Factors | | | | | | | | | |
| BruteForceMatcher | vs Baseline Matcher | BruteForce-106.58x | 31.65x | 43.71x | 42.08x | 52.91x | 59.39x | 70.11x | 77.84x | 69.61x | 100.78x |
| | vs Baseline TreeMatcher | Stateful-299.68x | 223.55x | 298.42x | 351.61x | 638.74x | 1100.59x | 835.15x | 991.31x | 1346.13x | 1096.15x |
| StatefulTreeMatcher | vs Baseline Matcher | BruteForce-137.84x | 49.74x | 71.37x | 73.25x | 91.57x | 102.71x | 81.24x | 98.40x | 134.28x | 126.55x |
| | vs Baseline TreeMatcher | Stateful-387.58x | 351.34x | 487.22x | 612.00x | 1105.57x | 1903.36x | 967.72x | 1253.07x | 2596.75x | 1376.49x |
| WhileLazyMatcher | vs Baseline Matcher | BruteForce-214.50x | 90.40x | 98.61x | 157.74x | 175.24x | 168.17x | 195.45x | 182.54x | 277.78x | 207.33x |
| | vs Baseline TreeMatcher | Stateful-603.14x | 638.53x | 673.16x | 1317.92x | 2115.71x | 3116.33x | 2328.17x | 2324.60x | 5371.96x | 2255.08x |
| FilteringParallelMatcher | vs Baseline Matcher | BruteForce- 44.71x | 31.36x | 41.85x | 43.60x | 55.69x | 62.97x | 61.00x | 47.20x | 78.55x | 91.48x |
| | vs Baseline TreeMatcher | Stateful-125.72x | 221.51x | 285.68x | 364.27x | 672.31x | 1166.99x | 726.56x | 601.08x | 1519.10x | 994.98x |
| LazyParallelMatcher | vs Baseline Matcher | BruteForce- 41.55x | 21.35x | 32.02x | 36.66x | 48.16x | 46.60x | 51.23x | 60.30x | 63.06x | 74.89x |
| | vs Baseline TreeMatcher | Stateful-116.83x | 150.77x | 218.61x | 306.27x | 581.40x | 863.63x | 610.18x | 767.95x | 1219.46x | 814.60x |



## C  Prototype Actor Implementations

Listing 12 shows the prototype actor implementations provided in the JoinActors library.

**Listing 12**  Actor implementation used in JoinActors: **Actor** (with support for join pattern matching) and **SimpleActor** (without join pattern matching support).

```scala
enum Result[+T]:
  case Stop(value: T)
  case Continue

import Result.*

class Actor[M, T](private val matcher: Matcher[M, Result[T]]):
  private val mailbox: Mailbox[M] = Mailbox[M]
  private val self           = ActorRef(mailbox)

  def start(): (Future[T], ActorRef[M]) =
    val promise = Promise[T]

    ec.execute(() => run(promise))

    (promise.future, self)

  @tailrec
  private def run(promise: Promise[T]): Unit =
    matcher(mailbox)(self) match
      case Continue    => run(promise)
      case Stop(value) => promise.success(value)

class SimpleActor[M, T](private val f: ActorRef[M] => PartialFunction[Any, Result[T]]):

  private val mailbox: Mailbox[M] = Mailbox[M]
  private val self = ActorRef(mailbox)

  def start(): (Future[T], ActorRef[M]) =
    val promise = Promise[T]

    ec.execute(() => run(promise))

    (promise.future, self)

  @tailrec
  private def run(promise: Promise[T]): Unit =
    f(self).applyOrElse[M, Result[T]](mailbox.take(), _ => Continue) match
      case Continue => run(promise)
      case Stop(value) => promise.success(value)
```

**About the authors**

**Ayman Hussein** (ayhu@dtu.dk) is a PhD student at DTU Compute (Technical University of Denmark).
https://orcid.org/0009-0005-6173-0976

**Philipp Haller** (phaller@kth.se) is an Associate Professor at KTH Royal Institute of Technology in Stockholm, Sweden.
https://orcid.org/0000-0002-2659-5271

**Ioannis Karras** (yaniskarras@gmail.com) is MSc in Computer Science and Engineering from DTU Compute (Technical University of Denmark).
https://orcid.org/0009-0006-6920-111X

**Hernán Melgratti** (hmelgra@dc.uba.ar) is an Adjunct Professor at the Universidad de Buenos Aires and a researcher at the Instituto de Investigación en Ciencias de la Computación (ICC), part of CONICET and the University of Buenos Aires (UBA).
https://orcid.org/0000-0003-0760-0618

**Alceste Scalas** (alcsc@dtu.dk) is an Associate Professor at DTU Compute (Technical University of Denmark).
https://orcid.org/0000-0002-1153-6164

**Emilio Tuosto** (emilio.tuosto@gssi.it) is a full professor in Computer Science at the Gran Sasso Science Institute (GSSI, L'Aquila - Italy).
https://orcid.org/0000-0002-7032-3281